\documentclass[aps, superscriptaddress, preprint, singlecolumn, longbibliography, showpacs, preprintnumbers,
amssymb, floatfix]{revtex4-1}

\usepackage{graphicx}
\usepackage{epsfig}
\usepackage{amsmath}
\usepackage{amssymb}
\usepackage{amsbsy}
\usepackage{xcolor}
\usepackage{soul}
\usepackage{hyperref}
\usepackage{amsthm}
\usepackage{nicefrac}

\usepackage{setspace}
\usepackage{esint}

\makeatletter
\usepackage{hyperref}

\usepackage{mathtools}

\DeclarePairedDelimiterX\braket[2]{\langle}{\rangle}{#1 \delimsize\vert #2}

\newcommand\ci{\mathrm{i}}

\newcommand\xunit{{\mbox{\boldmath{$\hat{x}$}}}}
\newcommand\yunit{{\mbox{\boldmath{$\hat{y}$}}}}

\newcommand\rhounit{{\mbox{\boldmath{$\hat{\rho}$}}}}
\newcommand\thetaunit{{\mbox{\boldmath{$\hat{\theta}$}}}}

\newcommand\kxunit{{\mbox{\boldmath{$\hat{k}_x$}}}}
\newcommand\kyunit{{\mbox{\boldmath{$\hat{k}_y$}}}}
\newcommand\krhounit{{\mbox{\boldmath{$\hat{k}_{\rho}$}}}}
\newcommand\kthetaunit{{\mbox{\boldmath{$\hat{k}_{\theta}$}}}}
\newcommand\kphiunit{{\mbox{\boldmath{$\hat{k}_{\phi}$}}}}

\begin{document}

\title{Building blocks for space-time non-separable pulses}
  \date{\today }
  

\author{Apostolos Zdagkas}
\affiliation{Optoelectronics Research Centre and Centre for Photonic Metamaterials, University of Southampton, Southampton SO17 1BJ, United Kingdom}
\author{Nikitas Papasimakis}
\affiliation{Optoelectronics Research Centre and Centre for Photonic Metamaterials, University of Southampton, Southampton SO17 1BJ, United Kingdom}
\author{Vassili Savinov}
\affiliation{Optoelectronics Research Centre and Centre for Photonic Metamaterials, University of Southampton, Southampton SO17 1BJ, United Kingdom}
\author{Nikolay I. Zheludev}
\affiliation{Optoelectronics Research Centre and Centre for Photonic Metamaterials, University of Southampton, Southampton SO17 1BJ, United Kingdom}
\affiliation{Centre for Disruptive Photonic Technologies, School of Physical and Mathematical Sciences and The Photonics Institute,
            Nanyang Technological University, Singapore 637378, Singapore}

\begin{abstract}
Space-time non-separable pulses hold promise for topological information transfer, probing ultra-fast light-matter interactions and engaging toroidal excitations in matter. Spurred by recent advances in ultra-fast and topological optics, these exotic electromagnetic excitations are now becoming the focus of growing experimental efforts. Many practical questions are yet to be answered regarding their generation, detection and light matter interactions. In particular, can these pulses be constructed from plane waves or other simple but experimentally accessible waves? Here we demonstrate that they can and as an example we provide analytical expressions for the characteristic case of the flying doughnut pulse which allows the expansion and synthesis of the pulse from sets of monochromatic beams, single-cycle pulses, and plane waves.
\end{abstract}
\maketitle

\section{Introduction}
The large majority of familiar electromagnetic waveforms (plane waves, Gaussian beams) are space-time separable, meaning that their spatial dependence can be separated from the temporal one. Such separation is however impossible for the more general family of space-time non-separable waves, which is far less know but far more common. Space-time non-separable light commonly arises as a result of non trivial light matter interactions or the diffraction and tight focusing of ultra-short pulses. This non-separability is key in understanding light matter interactions with ultra-short pulses and has important implications for topological optics \cite{zdagkas2019singularities}. Moreover, space-time non-separable pulses have emerged as a key component of toroidal electrodynamics in particular in the context of non-radiating configurations \cite{Raybould_anapole_excitation}. Here we focus on a characteristic example of a space-time coupled pulse termed Flying Doughnut (FD).

FD pulses were discovered in the context of early works on wide band and non-diffracting pulses  \cite{Localized_Waves_Ch_1}. In particular, with the advent of ultrashort lasers significant theoretical efforts were directed to the description of few-cycle pulses. In 1983, Brittingham proposed a localized solution to Maxwell's equations (focus wave modes, (FWMs)), where he also set the requirements for the mathematical formulation of this type of pulses \cite{Brittingham1983}. Although the derived solutions required infinite energy \cite{Comment_on_Brittigham1983}, Brittingham's efforts provided the basis for the study of non-diffracting pulses \cite{Localized_Waves_Ch_1}. Soon after, R. W. Ziolkowski showed that Brittingham's FWMs are modified Gaussian pulses that arise as solutions to the scalar wave equation with moving complex sources \cite{ziolkowski1985exact} and proposed that a superposition of such pulses, can be used to construct finite energy pulses termed ``electromagnetic directed-energy pulse trains'' (EDEPTs) \cite{ziolkowski1989localized}. Special cases of Ziolkowski's solutions were studied by Hellwarth and Nouchi, who found closed form expressions that describe single cycle pulses which are space-time non-separable finite energy solutions to the homogeneous Maxwell's equations. They showed that this family of pulses includes both linearly polarized pulses, termed ``pancakes'' \cite{Feng:98,PhysRevE.59.4630}, as well as pulses of toroidal symmetry, termed``focused doughnuts'' (FDs) \cite{PhysRevE.54.889}.

FDs are single-cycle pulses with toroidal electromagnetic field configuration and unusual spatio-temporal coupling \cite{PhysRevE.54.889}. FD pulses exhibit a fine topological structure along with regions where energy back-propagation occurs \cite{zdagkas2019singularities}, while even their interaction with simple homogeneous media is non-trivial \cite{Raybould_interaction}. They form the free-space propagating counterparts of the recently established toroidal excitations in matter, i.e. charge-current configurations with doughnut-like topology \cite{papasimakis2016toroidal}. In fact, it has been shown that flying doughnut pulses can efficiently engage dynamic toroidal and anapole excitations in dielectric particles, even when the latter does not posses toroidal symmetry \cite{Raybould_interaction,Raybould_anapole_excitation}.

Recent works have shown that FD pulses could be generated in the optical part of the spectrum from a laser pulse interacting with appropriately patterned metasurfaces \cite{papasimakis_generation, zdagkas2019generation}. Here, the generation of FD pulses requires to address simultaneously the few cycle nature, toroidal symmetry and importantly space-time coupling (STC). STCs appear as non-uniform spectral or temporal properties across a plane normal to the propagation vector of a pulse.  Although apparent in the time domain, STCs can be more intuitively studied in the frequency domain. Evidently, a plane wave expansion of the pulse is highly desirable for the generation and study of light-matter interactions of FD pulses.

In this work, we provide closed form expressions for the time-frequency Fourier transform of the pulse. We use the analytical expressions to show that the pulse does not exhibit STCs related to the spectral phase other than a pulse front curvature that is required for any diffracting pulse. The pulse has only STCs that are related to its spectral intensity. The latter render the pulse isodiffracting, a trait that renders the pulse resistant to changes of its shape and intensity upon propagation. We also present an expression for the spatial Hankel transform and provide a semi-analytical plane wave expansion of the FD pulse, where the role of back-propagating plane waves in the non-paraxial regime becomes clear. Finally, we introduce a new type of single-cycle plane-wave like electromagnetic excitations, the slice pulses, and demonstrate that this new set of linearly polarized broad-band solutions to Maxwell's equations can be used to decompose the Flying Doughnut pulses.

\section{Time-frequency Fourier transform of the flying doughnut}
\label{section:time_frequency}

\begin{figure}[t]
\centering
\includegraphics[width=1.0\linewidth]{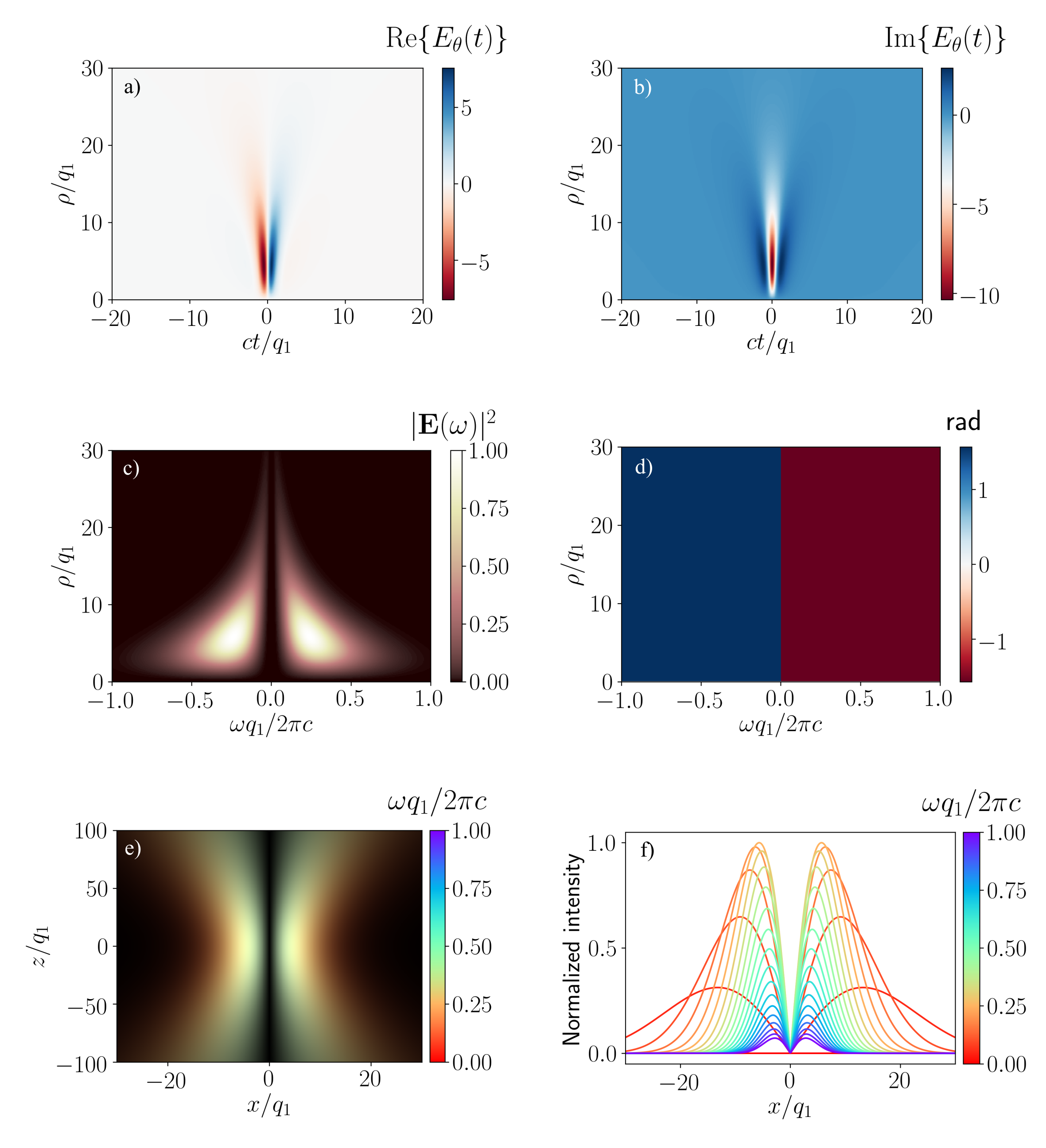}
\caption{a), b) Sections of the electric field ``$1 \frac{1}{2}$-cycle" and ``$1$-cycle" TE pulses respectively at their focus. c) Spectral intensity and d) phase of the TE ``$1 \frac{1}{2}$-cycle" pulse. e) False colour representation of an $x-z$ section of the electric field intensity of the beams for $q_2=100q_1$ at $y/q_1=0$ and for a number frequencies spanning the full bandwidth of the pulse's spectrum. The brightness represents the intensity distribution in space while the colour corresponds to the different wavelengths. Blue and red present shorter and longer wavelengths respectively. f) Normalized intensity of the radial distribution of the spectrum for each of the monochromatic beams. At the central part of the pulse all the wavelengths are present with the smaller being dominant. That leads to white-bluish colouring of the false colour image (see e)). At the edges only bigger wavelengths appear and hence the false colour image is coloured red. The existing space-time coupling is the consequence of the isodiffracting nature of the pulse.}
\label{Fig:fourier}
\end{figure}

The transverse electric (TE) Flying Doughnut pulse is given in the time-domain by the following equations
\begin{align}
E_{\theta} &=-4 \ci f_0 \sqrt{\frac{\mu_0}{\epsilon_0}} \frac{\rho \left( q_1 +q_2 -2\ci c t \right)}{\left[ \rho^2 + (q_1+\ci \tau)(q_2-\ci \sigma) \right]^3} \label{Eq:Etheta} \\
H_{\rho} &=4 \ci f_0 \frac{\rho \left( q_2 - q_1 -2\ci z \right)}{\left[ \rho^2 + (q_1+\ci \tau)(q_2-\ci \sigma) \right]^3}  \label{Eq:Hrho} \\
H_z &=-4 f_0  \frac{ \rho^2 - (q_1 + \ci \tau)(q_2 -\ci \sigma)}{\left[ \rho^2 + (q_1+\ci \tau)(q_2-\ci \sigma) \right]^3},  \label{Eq:Hz}
\end{align}
where $\epsilon_0$ and $\mu_0$ are the vacuum permittivity and permeability respectively, $\tau = z-ct$, $\sigma = z+ct$, $c$ is the speed of light in vacuum and $f_0$ a constant defining the amplitude and the units. The pulse is defined by $q_1$ and $q_2$ and must satisfy the inequality $q_1 \leq q_2$. When compared to a Gaussian beam, $q_1$ has the role of the wavelength and $q_2$ the role of the Rayleigh range. Both real and imaginary parts are solutions to Maxwell's equations describing ``$1 \frac{1}{2}$-cycle" and ``$1$-cycle" FD pulses respectively. The transverse magnetic (TM) pulse can be derived from Eq. \ref{Eq:Etheta}-\ref{Eq:Hz} by exchanging electric and magnetic fields
\begin{align}
\mathbf{E}_{\mathrm{TM}} &= \sqrt{\frac{\mu_0}{\epsilon_0}} \mathbf{H}_{\mathrm{TE}} \\
\mathbf{H}_{\mathrm{TM}} &= -\sqrt{\frac{\epsilon_0}{\mu_0}} \mathbf{E}_{\mathrm{TE}}.
\end{align}
and describes a radially polarized FD. Sections of the ``$1 \frac{1}{2}$-cycle" and ``$1$-cycle" TE pulses respectively at their focus are shown in Fig. \ref{Fig:fourier} a) and b). Although the name of the pulses seems contradictory, we should note that they were named from their far field appearance and not from their shape at focus.

In the frequency domain, the expression for the TE ``$1 \frac{1}{2}$-cycle" pulse can be analytically derived by a Fourier transform as (see App. \ref{App:Fourier})
\begin{align}\label{Eq:fourier:real2}
E_{\mathrm{re}} (\omega) =  \left\{
\begin{array}{ll}
4f_0\sqrt{\mu_0/\epsilon_0}\left\{\ci\pi \frac{\omega}{c^2} \frac{\rho}{q_1} e^{-\frac{\omega(q_2+q_1)}{2c}} \frac{e^{\frac{\ci\omega}{2c} \sqrt{A}}
                          \left( 2c \ci + \omega \sqrt{A} \right) +
                          e^{-\frac{\ci\omega}{2c} \sqrt{A}}
                          \left( -2c \ci + \omega \sqrt{A} \right)}
                     {2 A^{3/2}} \right\}^* & \omega > 0 \\
4f_0\sqrt{\mu_0/\epsilon_0} \ci\pi \frac{\omega}{c^2} \frac{\rho}{q_1} e^{\frac{\omega(q_2+q_1)}{2c}} \frac{e^{-\frac{\omega\ci}{2c} \sqrt{A}}
                          \left( -2c \ci + \omega \sqrt{A} \right) +
                          e^{\frac{\omega\ci}{2c} \sqrt{A}}
                          \left( 2c \ci + \omega \sqrt{A} \right)}
                     {2 A^{3/2}}  & \omega < 0 \\
      0 & \omega=0 \\
\end{array}
\right.
\end{align}
\noindent with
\begin{align}\label{Eq:fourier:A}
A \equiv A (\rho,z,q_2) = -(-q_1+q_2-2\rho-2\ci z)(-q_1+q_2+2\rho-2\ci z).
\end{align}
Similar expressions can be derived for the magnetic fields for both the real and imaginary parts and for the transverse electric and magnetic pulses (see App. \ref{App:Fourier}).

Equation \ref{Eq:fourier:real2} allows us to examine the spectral phase of the FD pulse. At the focus, $z=0$, and for radii where the pulse has non-negligible energy, the quantity $A$ is negative, leading to a purely imaginary field in the frequency domain. The latter is illustrated in Fig. \ref{Fig:fourier} d), where the radial distribution of the spectral phase of the pulse is plotted. As a result, the spectral phase is flat, indicating that the pulse is transform limited. The phase difference of $180$ degrees between the positive and negative frequencies is a property of the Fourier transform of a real function. Importantly, the pulse has a flat spectral phase across different radii, which indicates that the pulse does not exhibit STC related to spectral phase, such as pulse front tilt, that would limit the duration and energy density of the pulse. Away from focus, $z \neq 0$, the spectral phase is a function of $\rho^2$ indicating that only even order STCs are allowed, which essentially correspond a pulse front curvature and is expected for diffracting pulses.

The presence of intensity-related STCs is illustrated  in Fig. \ref{Fig:fourier} c), where we plot the radial distribution of spectral intensity of the pulse at focus, $z=0$. It is apparent that the spectrum close to the centre of the pulse spans over a much broader frequency range, than at larger radii. On the other hand, higher frequencies are more tightly spatially confined than lower frequencies. While usually STCs deteriorate the desired characteristics of a pulse, such as the intensity spatial profile and duration \cite{STCs_review2010}, in the case of the FD pulse, STCs are necessary to guaranty a well-behaved and stable pulse. We will show here that it is this STC that makes the pulse isodiffracting and ensures that all frequency components of the pulse diffract at the same rate. Consequently, the isodiffracting nature of the FD pulse confines the radial intensity distribution to the smallest possible area upon propagation, as compared to any other pulse with the same waist at focus. Indeed, any pulse comprising a superposition of beams that do not diffract at the same rate will suffer from an unequal spreading of the energy across different frequencies, which ultimately leads to a reduction of the energy density upon propagation. In contrast, an isodiffracting pulse, such as the FD, will exhibit, by definition, the highest energy density possible.

Isodiffracting beams exhibit a wavelength independent Rayleigh length, a property that Feng et al. showed to be true for the ``pancake'' pulse \cite{PhysRevE.59.4630,PhysRevE.61.862, Feng:99}. In this case, longer wavelengths exhibit larger waist regions than shorter wavelengths at any propagation distance as a result of the diffraction limit. This is evident at the focus of FDs by simply plotting the intensity distribution for each wavelength, Fig. \ref{Fig:fourier} e) and f). Fig. \ref{Fig:fourier} e) shows a false colour image of the intensity of the beams. The brightness represents the intensity distribution in space while the colour corresponds to different wavelengths. As it is shown in Fig. \ref{Fig:fourier} f) the central part of the pulse is occupied by almost all the frequencies with the shorter wavelengths having their maxima there. That makes the false colour image to seem white-blueish there. At the edges only longer wavelengths carry non-negligible energy thus leading to a reddish colour.

\begin{figure}[t]
\centering
\includegraphics[width=1.0\linewidth]{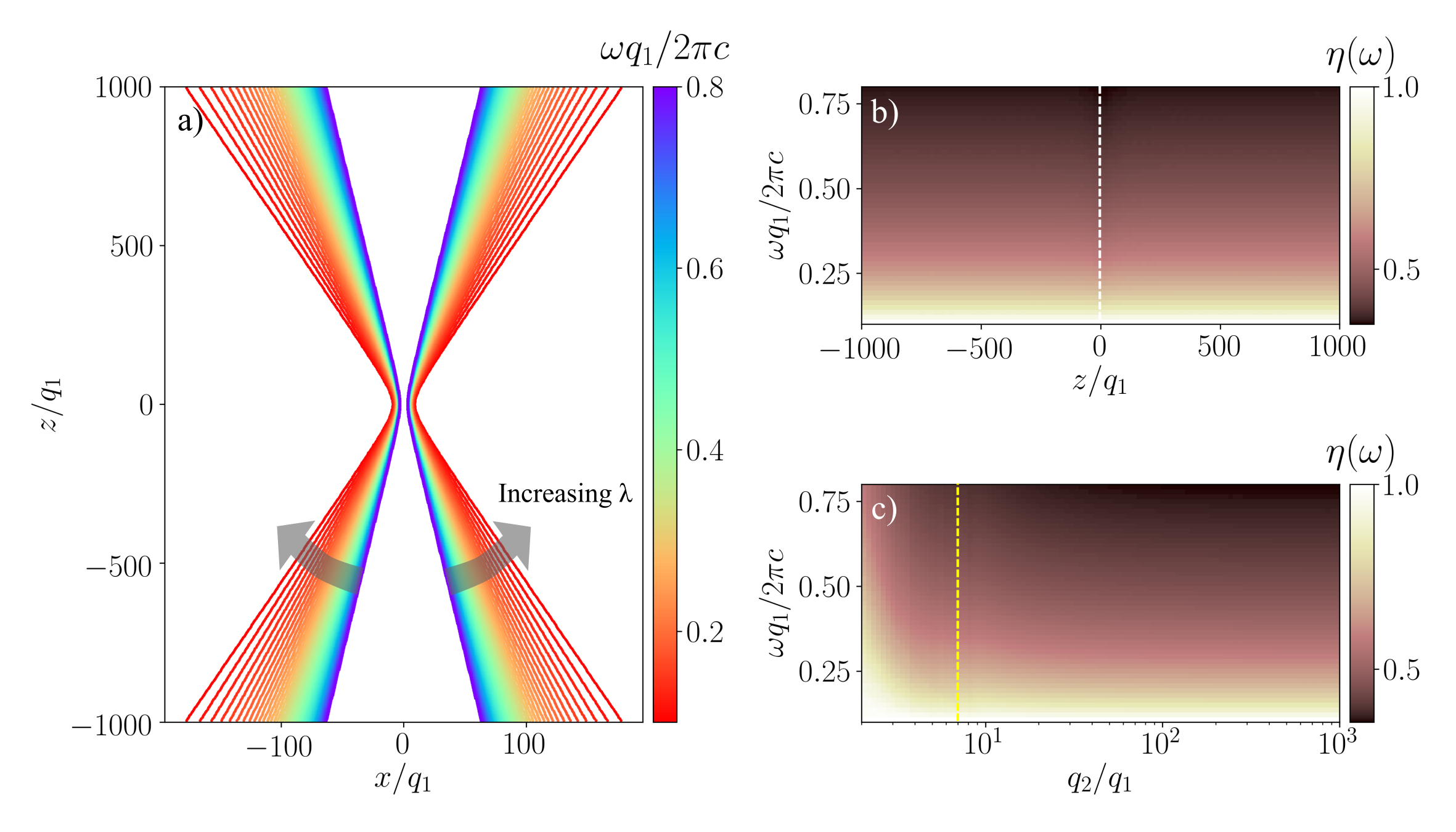}
\caption{a) Trace of the position of the maximum intensity for each wavelength. b) Numerical calculation of the ratio of the radial position of the maxima between a reference frequency, $\omega_1 q_1/2 \pi c  = 0.1$ and a variable frequency $\omega$, for $q_2=100q_1$ and for increasing propagation distance $z/q_1$. The white dashed line indicates the position of the focus. c) Numerical calculation of the same ratio for fixed distance, $z/q_1 = 1000$ and for increasing $q_2$. For $z \gg q_2$ and $q_2 \gg q_1$ the value of the ratio is similar to the theoretically predicted value from Eq. \ref{Eq:r_max_ratio}. The yellow dashed line indicates the limit of $q_2/q_1$ below which Eq. \ref{Eq:r_max_ratio} deviates from the actual value.}
\label{Fig:max_ratio}
\end{figure}

The isodiffraction property of the FD pulse can be rigorously derived in the case of well collimated pulses ($q_2 \gg q_1$), see App. \ref{App:isodiffraction}. In particular, the ratio of the intensity maxima of two different wavelengths can be analytically obtained as:
\begin{align}
\eta(\omega_2) = \frac{\rho_{\max} (\omega_2)}{\rho_{\max} (\omega_1)} = \sqrt{\frac{(4+(q_2^2-1)^2\omega_1^2)(-2+2\sqrt{5+(q_2^2-1)^2\omega_2^2})}{(-2+2\sqrt{5+(q_2^2-1)^2\omega_1^2})(4+(q_2^2-1)^2\omega_2^2)}}
\label{Eq:r_max_ratio}
\end{align}
and is independent of the propagation distance $z$.

To illustrate the validity of Eq. \ref{Eq:r_max_ratio}, we start by plotting the position of the maximum  intensity for each wavelength that is calculated numerically for increasing propagation distance $z/q_1$ and for $q_2=100q_1$, Fig. \ref{Fig:max_ratio} a). Then we use as a reference frequency the smallest frequency on the plot, $\omega_1 q_1/2 \pi c = 0.1$ and we calculate the ratio of the position of the maxima  for all the frequencies up to $\omega q_1/2 \pi c = 0.8$. We do this for various values of propagation distance $z/q_1$, Fig. \ref{Fig:max_ratio} b), or the parameter $q_2$, Fig. \ref{Fig:max_ratio} c). For $z \gg q_2$ and $q_2 \gg q_1$ the value of the ratio is similar to the theoretically predicted value from Eq. \ref{Eq:r_max_ratio} verifying its validity. Additionally we notice that the ratio is always smaller than $1$ which means that the shorter wavelengths have its maximum spectral intensity at smaller radii as it was expected. From Fig. \ref{Fig:max_ratio} b) we see that apart from propagation distances close to the focus, the ratio is independent of the propagation distance as it is predicted from Eq. \ref{Eq:r_max_ratio}. Finally, from Fig. \ref{Fig:max_ratio} c) we can see that for $q_2 \gg q_1$ the ratio becomes independent of the parameter $q_2$, meaning that for very well collimated beams the ratio of the position of the maxima between two frequencies can be described by a very simple expression that is proportional to the ratio of the frequencies.

\section{Spatial Hankel transform of the flying doughnut-Plane wave expansion}

The Hankel transform of the electric field for the TE pulse can be derived as (see App. \ref{App:hankel})
\begin{align}
\mathbf{E}_{\theta} (k_{\rho}) &= - 8 \pi f_0 \sqrt{\frac{\mu_0}{\epsilon_0}} (q_1 + q_2 -2 \ci ct) \frac{k^2_{\rho}}{8a} K_{1} (k_{\rho} \alpha) \thetaunit,
\label{Eq:hankel}
\end{align}
with $\alpha = \sqrt{(q_1 +\ci z -\ci ct)(q_2 -\ci z - \ci ct)}$ and $K_1$ the first order modified Bessel function of the second kind. The transforms of the real and imaginary parts, or the ``$1 \frac{1}{2}$-cycle" and ``$1$-cycle", are given by
\begin{align}
E_{\mathrm{re}, \theta} (k_{\rho}) &= \frac{ E_{\theta} (k_{\rho}) -  E^*_{\theta} (k_{\rho})}{2}
\label{Eq:hankel_re} \\
E_{\mathrm{im}, \theta} (k_{\rho}) &= \frac{ E_{\theta} (k_{\rho}) +  E^*_{\theta} (k_{\rho})}{2 \ci},
\label{Eq:hankel_im}
\end{align}
respectively.

\begin{figure}[t]
\centering
\includegraphics[width=1.0\linewidth]{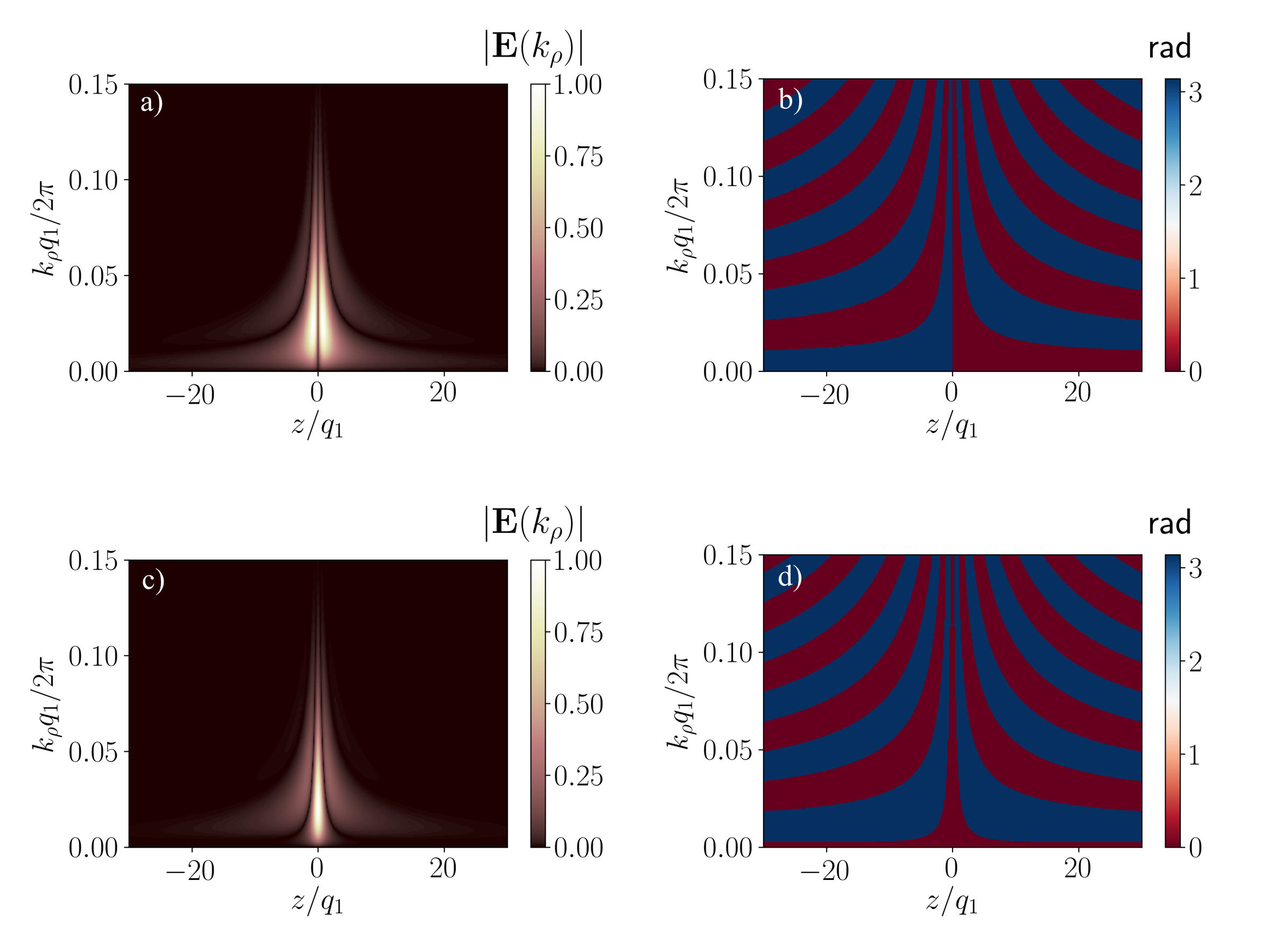}
\caption{Amplitude and phase of the Hankel transforms of the real, a) and b), and imaginary, c) and d), electric fields at focus, $t=0$, and for $q_2=100q_1$.}
\label{Fig:hankel}
\end{figure}

Figure \ref{Fig:hankel} shows the electric field Hankel transforms of the real $1 \frac{1}{2}$-cycle (Fig. \ref{Fig:hankel}a-b), and imaginary $1$-cycle (Fig. \ref{Fig:hankel}c-d) pulses at focus ($t=0$) for $q_2=100q_1$. The former is antisymmetric with respect to the $z$ coordinate and vanishes at $z=0$ and $t=0$ (see Fig. \ref{Fig:fourier}a). On the other hand, the Hankel transform of the $1$-cycle pulse is symmetric with respect to $z$ and peaks at $z=0$. Similarly to the time-frequency Fourier transform case, the Hankel transform exhibits a flat radial spectral phase profile at $z=0$ (Fig. \ref{Fig:hankel}b\& d) resulting in a radially transform-limited pulse. The latter indicates that given a focusing strength the constituting pulses as described in the spatial frequency domain are all in-phase. Such radially transform limited pulses have the smallest possible focal spot, which is controlled only by the radial spectral bandwidth. Moreover, the sign of the phase alternates with increasing spatial frequency, which is related to the isodiffracting nature and fine topological structure of the pulse \cite{zdagkas2019singularities}.

\begin{figure}[t]
\centering
\includegraphics[width=1.0\linewidth]{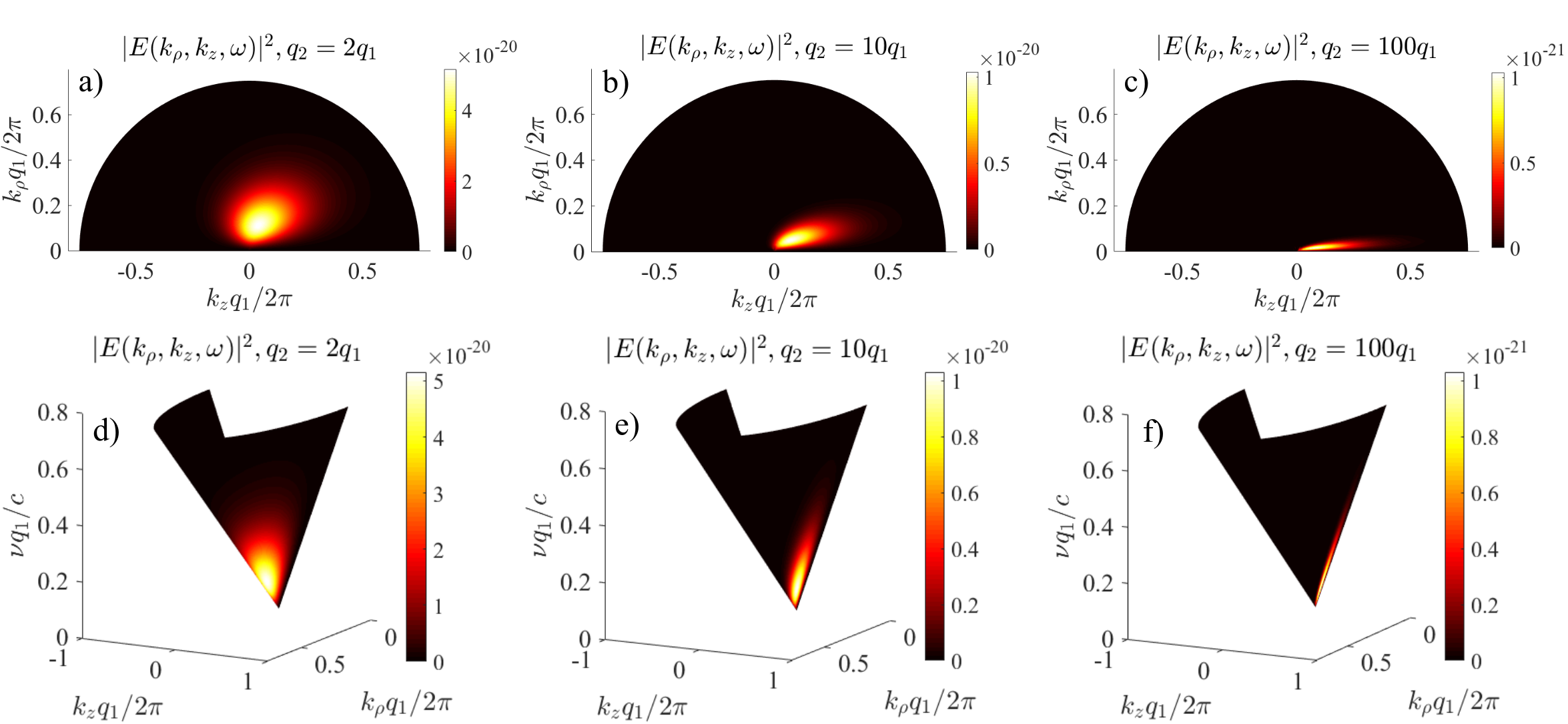}
\caption{Plane wave decomposition of the FD pulse as it is presented on the surface of a half-cone, representing the cone of light. a)-c) The $k_\rho-k_z$ projection of the decomposition. d)-f) Full three dimensional presentation of the cone. The colours represent the squared amplitude of the plane waves. The plot is given only for positive frequencies.}
\label{Fig:light_cone}
\end{figure}

The  Hankel transform allows to readily expand the FD pulse into plane waves. Indeed, for the plane wave expansion of the FD pulse, two spatial ($(\rho,z)\rightarrow (k_{\rho},k_z)$) and one temporal ($t\rightarrow\omega$) transforms are required, of which only one at a time can be derived analytically (either time-frequency Fourier or Hankel transform). Since the radial transform is given by a simple analytical expression, the Fourier transforms of the two remaining dimensions, time and longitudinal position, can be easily performed numerically. The result of the plane wave decomposition is presented in Fig. \ref{Fig:light_cone} for three different cases, $q_2=2q_1$, $q_2=10q_1$ and $q_2=100q_1$. The decomposition is given at the surface of a cone that represents the cone of light. Only positive $k_\rho$ are presented since the pulse is azimuthally symmetric. The first row depicts the projection of that cone to the $k_\rho-k_z$ plane while in the second row a 3D image of the cone is presented. The colours represent the squared amplitude of the plane waves. The plot is given only for positive frequencies. Figure \ref{Fig:light_cone} indicates that in the extreme non-paraxial regime, where $q_2 \simeq q_1$ backward propagating waves are present in the FD pulse. This is revealed by the appearance of negative $k_z$ for positive frequencies. Now by increasing $q_2$ with respect to $q_1$ we notice a reduction of $k_\rho$ in favour of $k_z$. This renders the pulse weakly focused and hence it propagates as a paraxial pulse ($q_2 \gg q_1$). Here, the intensity STC reveals itself as a narrower bandwidth for small radial wavevectors $k_\rho$ (see also discussion in Sec. \ref{section:time_frequency}).

\section{Decomposition of FD pulses into linearly polarized single-cycle pulses}

Overall, the plane wave decomposition can at a glance fully characterize the pulse and provide insights into its propagation properties. However, a more natural choice of basis for such short cycle pulses would be a set of functions that themselves are single-cycle. Here we present such a set of single-cycle functions, which we term the `slice pulses', and demonstrate that any FD pulse can be fully represented in terms of these slice pulses.
 
The decomposition of the Flying Doughnut pulse into the Slice pulses is based on the following conjecture: FD pulse consists exclusively out of the luminal plane waves, i.e. plane waves that propagate at the speed of light. It follows therefore that if one was to apply a `luminal' filter, to the FD pulse, that would suppress all the non-luminal plane-wave components of the FD pulse, the filtered pulse would be precisely the initial FD pulse.     

Define, non-luminal plane waves as those for which ratio of the wavenumber ($k$) and angular frequency ($\omega$) does not reduce to speed of light ($c$), i.e. $\omega/k\neq c$. The `luminal filter' that would remove the non-luminal plane waves from any electromagnetic
pulse can be imagined as an on-off function in the frequency space,
i.e.:
\[
\tilde{h}\left(\omega,\,\vec{k}\right)=\lim_{\xi\to0}\begin{cases}
1, & -\xi<\left(c^{2}\left|k\right|^{2}-\omega^{2}\right)<\xi\\
0, & otherwise
\end{cases}
\]

Where $\vec{k}$ is the wave-vector. Multiplying the Fourier-transformed
electromagnetic field by such a filter would remove all non-luminal
plane-wave components. In the Appendix (see App.~\ref{app:SpaceTimeFilter}),
it is shown that in spacetime domain such a filter is given by:
\begin{multline}
h\left(t-t',\,\left|\vec{r}-\vec{r}'\right|\right)=\frac{-i}{4\pi^{2}}\lim_{\epsilon\to0}\int_{-\frac{\epsilon}{2}}^{\frac{\epsilon}{2}}d\zeta\left(\zeta-\frac{\partial_{t}}{ic}\right)\Biggl\{\\
\frac{\delta\left(\frac{\left|\vec{r}-\vec{r}'\right|}{c}-t+t'\right)\exp\left(i\zeta\left|\vec{r}-\vec{r}'\right|\right)-\delta\left(\frac{\left|\vec{r}-\vec{r}'\right|}{c}+t-t'\right)\exp\left(-i\zeta\left|\vec{r}-\vec{r}'\right|\right)}{\left|\vec{r}-\vec{r}'\right|}\Biggr\}\label{eq:MailFilterEq}
\end{multline}

The luminal filter is a scalar function. Applying it to vector-valued
electromagnetic fields of the FD pulse is ambiguous. However, the
FD pulse itself arises from the function, which we shall term the
`seed-function':
\begin{equation}
f\left(t,\,\vec{r}\right)=f\left(t,\,\rho,z\right)=\frac{1}{\rho^{2}-\left(\left(ct-z\right)+iq_{1}\right)\left(\left(ct+z\right)+iq_{2}\right)}\label{eq:MainSeedDef}
\end{equation}

The electric ($\vec{E}$) and magnetic ($\vec{B}$) fields of the
TE FD pulse are then \cite{Hellwarth96}:
\begin{gather*}
\vec{E}=-\partial_{t}\boldsymbol{\nabla}\times\vec{\hat{z}}f,\quad\vec{B}=\boldsymbol{\nabla}\times\boldsymbol{\nabla}\times\vec{\hat{z}}f
\end{gather*}

Since plane-waves are eigen-functions of the above derivatives, to
filter out the non-luminal plane wave components of the FD pulse it
is sufficient to apply the filter to the seed function itself. In
the appendix (App.~\ref{app:FDSeed}) it is shown that the convolution
of the seed function with the luminal filter results in:
\begin{flalign*}
f=h\otimes f= & -\frac{1}{\pi}\oint d^{2}\Omega'\cdot\mathcal{S}\left(t,\,\vec{r};\,\vec{\hat{r}}'\right)\\
\mathcal{S}\left(t,\,\vec{r};\,\vec{\hat{r}}'\right)= & \frac{\left[2\vec{r}.\vec{\hat{r}}'+i\left(\vec{\hat{z}}.\vec{\hat{r}}'\right)\left(q_{2}-q_{1}\right)\right]^{2}+\left[2ct+i\left(q_{2}+q_{1}\right)\right]^{2}}{\left(\left[2\vec{r}.\vec{\hat{r}}'+i\left(\vec{\hat{z}}.\vec{\hat{r}}'\right)\left(q_{2}-q_{1}\right)\right]^{2}-\left[2ct+i\left(q_{2}+q_{1}\right)\right]^{2}\right)^{2}}
\end{flalign*}

Thus the Flying Doughnut seed function can be represented in terms
of an integral over the solid angle of a different function ($\mathcal{S}\left(t,\,\vec{r};\,\vec{\hat{r}}'\right)$),
which we term the `slice function'. Crucially, as the seed function,
the slice function can be shown to be a solution of the scalar wave
equation, i.e. $\partial_{tt}\mathcal{S}-c^{2}\nabla^{2}\mathcal{S}=0$.
It follows that the electromagnetic fields of the Flying Doughnut
pulses can be thought of as being linear superpositions of a different
family of `slice pulses', themselves solutions to Maxwell's equations,
parameterized by unit-vector $\vec{\hat{r}}'$:
\begin{gather*}
\vec{E}_{FD}=-\frac{1}{\pi}\oint d^{2}\Omega'\,\vec{E}_{\mathcal{S}}\left(t,\vec{r};\vec{\hat{r}}'\right),\quad\vec{B}_{FD}=-\frac{1}{\pi}\oint d^{2}\Omega'\,\vec{B}_{\mathcal{S}}\left(t,\vec{r};\vec{\hat{r}}'\right)\\
\vec{E}_{\mathcal{S}}=-\partial_{t}\boldsymbol{\nabla}\times\vec{\hat{z}}\mathcal{S},\quad\vec{B}_{\mathcal{S}}=\boldsymbol{\nabla}\times\boldsymbol{\nabla}\times\vec{\hat{z}}\mathcal{S}
\end{gather*}

Silce function can be written in a compact form as follows:
\begin{gather}
\mathcal{S}=\mathcal{S}_{0}\cdot\frac{\mathcal{A}^{2}+\mathcal{B}^{2}}{\left(\mathcal{A}^{2}-\mathcal{B}^{2}\right)^{2}}\label{eq:slice_gather}
\\
\mathcal{A}=2\vec{r}.\vec{\hat{r}}'+i\left(\vec{\hat{z}}.\vec{\hat{r}}'\right)\left(q_{2}-q_{1}\right),\quad\mathcal{B}=2ct+i\left(q_{2}+q_{1}\right)\label{eq:slice_AB}
\end{gather}

The electric field of a slice pulse is:
\begin{equation}
\vec{E}_{\mathcal{S}}=E_{0}\cdot\frac{\mathcal{AB}}{\left(\mathcal{A}^{2}-\mathcal{B}^{2}\right)^{3}}\cdot\left(\vec{\hat{r}'}\times\vec{\hat{z}}\right),\quad E_{0}=96c\mathcal{S}_{0}\label{eq:EfieldSlice}
\end{equation}

Several properties of the slice pulses can now be noted. Firstly, the slice pulses, unlike the Flying Doughnuts, are linearly polarized.
The spatial dependence is fully contained in $\mathcal{A}$ which is linear in $\vec{r}.\vec{\hat{r}'}$. It follows that slice pulses are constant in planes that are perpendicular to $\vec{\hat{r}}'$. Next one can consider the speed of propagation of these pulses. The ratio of the spatial and temporal derivatives of the slice function is:

\begin{equation}
\frac{\vec{\nabla}\mathcal{S}}{\partial_{ct} \mathcal{S}} = \vec{\hat{r}'} \cdot \frac{\mathcal{A}}{\mathcal{B}}\cdot \frac{\left(\mathcal{A}^2-\mathcal{B}^2\right)^2 - \left(\mathcal{A}^4-\mathcal{B}^4\right)}{\left(\mathcal{A}^2-\mathcal{B}^2\right)^2 + \left(\mathcal{A}^4-\mathcal{B}^4\right) }   
\end{equation}

Clearly (see Eq.~(\ref{eq:slice_gather})) the peak in the slice function is at $\mathcal{A}\approx\mathcal{B}$, which corresponds to ratio

\[
\frac{\vec{\nabla}\mathcal{S}}{\partial_{ct} \mathcal{S}} \approx \vec{\hat{r}'} \cdot \frac{\mathcal{A}}{\mathcal{B}}\cdot \left(-1\right)  \approx -\vec{\hat{r}'} 
\]

which is consistent with propagation with constant speed, the speed of light, along the $\vec{\hat{r}'}$-direction. Finally we consider the amplitude of the slice pulse in the $\vec{\hat{r}'}$-direction. Since $\mathcal{A}$ is linear in $\vec{\hat{r}'}$ (see Eq.~(\ref{eq:slice_AB})), one can easily establish that the electric field of the slice pulse decays as $E\propto 1/\left(\vec{\hat{r}'}.\vec{r}\right)^5$ (see Eq.~(\ref{eq:EfieldSlice})), i.e. slice pulses are localized in the propagation direction.

In summary slice pulses are linearly polarized transverse pulses with infinite extent and constant value over each plane perpendicular to $\vec{\hat{r}'}$, but strong localization along the direction $\vec{\hat{r}'}$. These pulses propagate at the speed of light along $\vec{\hat{r}'}$-direction. A nice way to visualize slice pulses is short sections of plane waves, truncated along the direction of propagation in such a way as to keep the resultant pulse a valid solution of the Maxwell's equations.

\begin{figure}[t]
\centering
\includegraphics[width=0.5\linewidth]{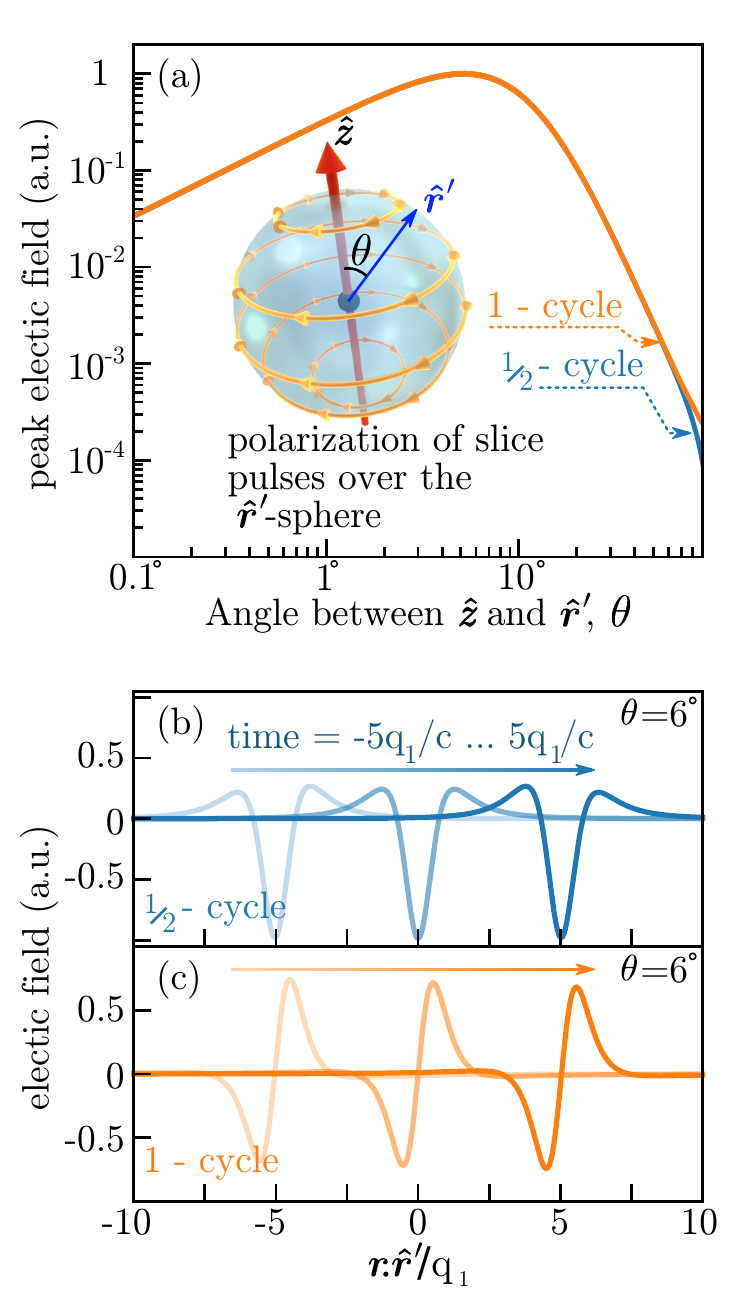}
\caption{The decomposition of the flying doughnut (FD) pulse with $q_2=100 q_1$ into the slice pulses. Each slice pulse is parametrized by unit-vector $\boldsymbol{\hat{r}'}$, and the FD pulse is obtained by integrating all the slice pulses over the surface of an $\boldsymbol{\hat{r}'}$-sphere. All slices depend on time $t$, and on projection of position-vector ($\boldsymbol{r}$) along the parameterization unit vector ($\boldsymbol{r.{\hat{r}'}}$). The polarization of (electric field) for all slice pulses is linear and $\propto \boldsymbol{\hat{r}'\times \hat{z}}$, where $\boldsymbol{\hat{z}}$ is the direction of propagation of the FD pulse. (a) Shows the maximum amplitude of the single-cycle (1-cycle) and half-cycle ($\nicefrac{1}{2}$-cycle) as a function of angle between the $\boldsymbol{\hat{z}}$ and $\boldsymbol{\hat{r}}'$. The inset shows the polarization of all slice pulses as a function of $\boldsymbol{\hat{r}}'$ (i.e. on the surface of the unit sphere). (b) Shows the electric field of a half-cycle slice pulse as a function of position ($\boldsymbol{r}$) for three different times ($c$ is the speed of light). (c) Shows the electric field of a single-cycle slice pulse.}
\label{Fig:SphereFull}
\end{figure}

Figure~\ref{Fig:SphereFull} shows the decomposition of the FD pulse into slice pulses (see Eq.~(\ref{eq:EfieldSlice})). The amplitude of the electric field of the individual slice pulses (in the FD decomposition) as a function of the angle between the propagation axis of the FD pulse ($\vec{\hat{z}}$), and the propagation axis of the slice pulse ($\vec{\hat{r}'}$), shown in Fig.~\ref{Fig:SphereFull}a, demonstrates that the FD pulse is composed of the slice pulses propagating primarily along the propagation of the FD pulse. Figure~\ref{Fig:SphereFull}b.c show the magnitude of the electric field of the individual slice pulses along the direction of propagation. As with FD pulses, the slice pulses come in two forms: single-cycle and half-cycle, both propagating at the speed of light.

\section{Summary}
In this paper we have presented closed form expressions for the Fourier transform and provided a frequency domain description of the FD pulse. The Fourier transform expression was used to prove that the FD pulses exhibits only intensity STCs which ensure that the pulse is isodiffracting, which improves its stability upon propagation. In addition to the description and analysis of STCs, the Fourier decomposition of the pulse into monochromatic beams facilitates the study of its propagation properties. These monochromatic beams are non-paraxial solutions of Maxwell’s equations and enable the use of efficient frequency domain propagation techniques. Thus, such a decomposition can be used not only for the study of the pulse propagation dynamics, but also, for the description of the interaction of the FD pulse with matter. The Hankel transform derived here allows to describe the radial spectrum in momentum space and hence reveals information related to the focusing properties of the pulse. As a closed form expression it is a well behaved function in contrast to the numerical Hankel transform that is inaccurate close to the axis \cite{Norfolk:10}. As such, it can provide an accurate picture of the spatial frequency distribution of the pulse and hence it was used for the decomposition of FDs into plane waves. Finally, an alternative decomposition of the FD into linearly polarized single-cycle pulses was provided, which may allow to construct toroidal pulses with prescribed, more complex, polarization profiles.

 Overall, our work allows to describe and analyze an ideal FD pulse in terms of propagation stability and energy distribution, which would be important in all applications involving pulsed energy transfer, such as free space telecommunications or ultrafast machining. The complete spatio-temporal description presented in this work provides a framework for the generation, detection, and study of light-matter interactions of complex space-time non-separable pulses, such as the FD.

\begin{acknowledgments}
The authors acknowledge the support of the MOE Singapore (MOE2016-T3-1-006), the UK’s Engineering and Physical Sciences Research Council (grant EP/M009122/1, Funder Id: http://dx.doi.org/10.13039/501100000266), the European Research Council (Advanced grant FLEET-786851, Funder Id: http://dx.doi.org/10.13039/501100000781), and the Defense Advanced Research Projects Agency (DARPA) under the Nascent Light Matter Interactions program. The data from this paper can be obtained from the University of Southampton ePrints research repository: https://doi.org/10.5258/SOTON/XXXXX.
\end{acknowledgments}

\appendix

\section{Fourier transform}
\label{App:Fourier}
In this appendix a step by step derivation of an analytical expression for the time-frequency Fourier transform of the FD pulse is presented.
The following Fourier transform pair is used
\begin{align}
{\bf F} \left({\bf r},\omega \right) &= \int_{-\infty}^\infty \mathrm{e}^{\ci\omega t} {\bf F} \left({\bf r},t \right) dt \\
{\bf F} \left({\bf r},t \right) &= \frac{1}{2\pi} \int_{-\infty}^\infty  \mathrm{e}^{-\ci\omega t} {\bf F} \left({\bf r}, \omega  \right) d\omega.
\label{App:fourier:fourier_def}
\end{align}

For the transverse electric field (TE) we have \cite{PhysRevE.54.889}
\begin{align}
E_{\theta} &=-4 \ci f_0 \sqrt{\frac{\mu_0}{\epsilon_0}} \frac{\rho \left( q_1 +q_2 -2\ci c t \right)}{\left[ \rho^2 + (q_1+\ci \tau)(q_2-\ci \sigma) \right]^3} \label{App:fourier:Etheta} \\
H_{\rho} &=4 \ci f_0 \frac{\rho \left( q_2 - q_1 -2\ci z \right)}{\left[ \rho^2 + (q_1+\ci \tau)(q_2-\ci \sigma) \right]^3}  \label{App:fourier:Hrho} \\
H_z &=-4 f_0  \frac{ \rho^2 - (q_1 + \ci \tau)(q_2 -\ci \sigma)}{\left[ \rho^2 + (q_1+\ci \tau)(q_2-\ci \sigma) \right]^3},  \label{App:fourier:Hz}
\end{align}
with $\tau = z-ct$ and $\sigma = z+ct$. For convenience and generality, we will use dimensionless variables. More specifically, we write everything with respect to $q_1$ which has dimensions of length. As such, we define $\rho^{\prime} = \rho/q_1$, $z^{\prime} = z/q_1$, $q_2^{\prime} = q_2/q_1$, $t^{\prime} = ct/q_1$ and $\omega^{\prime} = q_1\omega/c$ and we omit the coefficients $4f_0\sqrt{\mu_0/\epsilon_0}/q_1^4$ and $4f_0/q_1^4$ for the electric and the magnetic fields respectively. Finally, we omit the primes on the new dimensionless variables for clarity. Now, the dimensionless fields are given by the following equations
\begin{align}
E_{\theta} &= -\ci \frac{\rho(1+q_2-2\ci t)}{[\rho^2 +(1+\ci z - \ci t)(q_2-\ci z -\ci t)]^3} \label{App:fourier:Ethetadimless} \\
H_{\rho} &= \ci \frac{\rho(q_2-1-2\ci z)}{[\rho^2 +(1+\ci z - \ci t)(q_2-\ci z -\ci t)]^3}
\label{App:fourier:Hrhodimless} \\
H_{z} &= - \frac{\rho^2-(1+\ci z - \ci t)(q_2-\ci z -\ci t)}{[\rho^2 +(1+\ci z - \ci t)(q_2-\ci z -\ci t)]^3}.
\label{App:fourier:Hzdimless}
\end{align}

We will first work with the electric field. From now on and for clarity we will refer to the electric field as $E$, but we actually mean that we are using the $\theta$ component. The real and imaginary parts of the field are quite complex expressions to compute the Fourier integral. Thus, and because of the linearity of the integral operator, we will calculate the Fourier transform of the complex field and then we will take the real and imaginary parts from the equations,
\begin{align}\label{App:fourier:real1}
E_{\mathrm{re}} (\omega) =\frac{E(\omega)+E^*(-\omega)}{2}
\end{align}
\noindent and
\begin{align}\label{App:fourier:im1}
E_{\mathrm{im}} (\omega) =\frac{E(\omega)-E^*(-\omega)}{2\ci}.
\end{align}
\begin{proof}
\begin{align*}
E_{\mathrm{re}} (\omega) &= \int_{-\infty}^{+\infty}  e^{\ci \omega t} \mathrm{Re} \left\{E(t)\right\} dt\\
&=\int_{-\infty}^{+\infty} \frac{e^{\ci \omega t} E(t) + [e^{-\ci \omega t} E(t)]^*}{2} dt \\
&=\frac{E(\omega)+E^*(-\omega)}{2}.
\end{align*}
\end{proof}
\noindent These correspond to the ``$1 \frac{1}{2}$-cycle" and ``$1$-cycle" respectively.

Returning now to the equation \ref{App:fourier:Ethetadimless}, it is apparent that we can apply Jordan's lemma since the power of $t$ on the denominator is 5 orders bigger than that of the numerator. That is the Fourier transform is given by the integral residues on the upper and lower half complex plane \cite{arfken2012mathematical}. One only has to find the poles and determine when they are located in upper half or lower half plane.

From \ref{App:fourier:Ethetadimless}, it is apparent that the equation has 2 triple poles, thus only two distinct. Luckily, they are both located in the lower half plane, though the algebra to prove this is elaborate. We are going to prove this explicitly, though a smart way to prove it can be found on \cite{PhysRevE.59.4630}.

We start by writing down the poles
\begin{align}
t_1 &= \frac{1}{2} \left( \sqrt{-(-1+q_2-2\rho-2\ci z)(-1+q_2+2\rho-2\ci z)} -\ci q_2-\ci \right)
 \label{App:fourier:pole1} \\
t_2 &= \frac{1}{2} \left( - \sqrt{-(-1+q_2-2\rho-2\ci z)(-1+q_2+2\rho-2\ci z)} -\ci q_2-\ci \right).
\label{App:fourier:pole2}
\end{align}
It is useful here to define
\begin{align}\label{App:fourier:A1}
A \equiv A (\rho,z,q_2) &= -(-1+q_2-2\rho-2\ci z)(-1+q_2+2\rho-2\ci z) \\ \nonumber
&=4z^2 + 4\rho^2 - (q_2-1)^2 + 4(q_2-1)z\ci.
\end{align}
Now we want to prove that
\begin{align}
\mathrm{Im} \left\{
\begin{array}{ll}
t_1 \\
t_2
\end{array}
 \right\} =&
 \frac{1}{2} \left\{ -1-q_2 \pm \left[ (4z-4zq_2)^2+(4\rho^2+4z^2-(q_2-1)^2)^2 \right]^{1/4} \right. \\ \nonumber
 &\left. \sin \left[ \frac{1}{2} \mathrm{Arg} (A) \right] \right\}
\end{align}
is negative for every $q_2 \geq 1, \rho \geq 0$ and $z$. We will prove it for $t_1$.
\begin{proof}
\begin{align}
\mathrm{Arg}(x+\ci y) = \mathrm{atan2}(y,x) = \left\{
\begin{array}{ll}
\arctan \left( \frac{y}{x} \right) & x >0 \\
\arctan \left( \frac{y}{x} \right) + \pi & x <0 \text{ and } y \geq 0 \\
\arctan \left( \frac{y}{x} \right) - \pi & x <0 \text{ and } y < 0 \\
+\frac{\pi}{2} & x =0 \text{ and } y > 0 \\
-\frac{\pi}{2} & x =0 \text{ and } y < 0 \\
\text{undefined} &  x=0 \text{ and } y=0.
\end{array}
\right.
\label{App:fourier:atan2}
\end{align}
\begin{itemize}
\item For $q_2=1$ or $z=0$, it is immediately apparent from \ref{App:fourier:pole1} that $\mathrm{Im} \left\{t_1 \right\} <0$.

\item For $z<0$, $\mathrm{Arg}(A)<0$ and thus $\mathrm{Im} \left\{t_1 \right\} <0$.

\item For $z>0$, we make use of the trigonometric identity $\sin(\arctan(x))=\frac{x}{1+x^2}$ and some other more common identities. It can be shown that all the remaining cases
\begin{itemize}
\item $4z^2+4\rho^2-(q_2-1^2)>0$
\item $4z^2+4\rho^2-(q_2-1^2)<0$
\item $4z^2+4\rho^2-(q_2-1^2)=0$
\end{itemize}
give $\mathrm{Im} \left\{t_1 \right\} <0$.
\end{itemize}
\end{proof}
\noindent A similar analysis holds for the second pole as well. Knowing that the poles are located in the lower half complex plane, the integral can be calculated from the integral residues
\begin{align}
\int_{-\infty}^{+\infty}  e^{\ci \omega t} E(t) dt = 2 \pi \ci \sum_i \mathrm{Res} \left\{e^{\ci \omega t}E(t),t_i\right\} I,
\end{align}
with $I$ denoting the sign of the contour (positive for anticlockwise). In general we have
\begin{align}
E(\omega) = -2 \pi \ci \left( \mathrm{Res} \left\{e^{\ci \omega t}E(t),t_1\right\} +
\mathrm{Res} \left\{e^{\ci \omega t}E(t),t_2\right\} \right),
\end{align}
and from \ref{App:fourier:real1}, the following cases arise
\begin{itemize}
\item $(\omega > 0)$
\begin{align}
&E(\omega) = 0 \\
&E^*(-\omega) = \left( -2 \pi \ci \left( \mathrm{Res} \left\{e^{-\ci \omega t}E(t),t_1\right\} +
\mathrm{Res} \left\{e^{-\ci \omega t}E(t),t_2\right\} \right) \right)^*
\end{align}
\item $(\omega < 0)$
\begin{align}
&E(\omega) = -2 \pi \ci \left( \mathrm{Res} \left\{e^{\ci \omega t}E(t),t_1\right\} +
\mathrm{Res} \left\{e^{\ci \omega t}E(t),t_2\right\} \right) \\
&E^*(-\omega) = 0.
\end{align}
\end{itemize}
There are no poles in the upper half plane, but reversal of the sign of $\omega$ is equivalent to integrating over the path of the lower half plane. It is advisable to calculate the  residues for higher order poles (triple in our case) using a computer algebra system (like Mathematica) in order to avoid mistakes in the trivial but error-prone procedure of computing the derivatives. Finally the residues are
\begin{align}
\mathrm{Res}\left\{ e^{\ci \omega t} E(t), t_1 \right\} &=
 -\frac{e^{\frac{\omega}{2}\left(1+q_2+\ci \sqrt{A} \right)}\omega\rho \left( 2\ci+\omega\sqrt{A} \right)}{2A^{3/2}} \\
 \mathrm{Res}\left\{ e^{\ci \omega t} E(t), t_2 \right\} &=
 -\frac{e^{\frac{\omega}{2}\left(1+q_2-\ci \sqrt{A} \right)}\omega\rho \left( -2\ci+\omega\sqrt{A} \right)}{2A^{3/2}} \\
\mathrm{Res}\left\{ e^{-\ci \omega t} E(t), t_1 \right\} &=
 -\frac{e^{-\frac{\omega}{2}\left(1+q_2+\ci \sqrt{A} \right)}\omega\rho \left( -2\ci+\omega\sqrt{A} \right)}{2A^{3/2}} \\
 \mathrm{Res}\left\{ e^{-\ci \omega t} E(t), t_2 \right\} &=
 -\frac{e^{-\frac{\omega}{2}\left(1+q_2-\ci \sqrt{A} \right)}\omega\rho \left( 2\ci+\omega\sqrt{A} \right)}{2A^{3/2}}
\end{align}

For $\omega=0$, it doesn't matter which contour we choose and knowing that we do not have poles in the upper half plane and the line of real values, the calculus of residues gives immediately the answer of having a zero integral. That is, there are no dc components in the field.

\indent Finally, the Fourier transform for the TE ``$1 \frac{1}{2}$-cycle" pulse is given by the equation
\begin{align}\label{App:fourier:real2}
E_{\mathrm{re}} (\omega) =  \left\{
\begin{array}{ll}
\left\{\ci\pi \omega\rho e^{-\frac{\omega(q_2+1)}{2}} \frac{e^{\frac{\ci\omega}{2} \sqrt{A}}
                          \left( 2 \ci + \omega \sqrt{A} \right) +
                          e^{-\frac{\ci\omega}{2} \sqrt{A}}
                          \left( -2 \ci + \omega \sqrt{A} \right)}
                     {2 A^{3/2}} \right\}^* & \omega > 0 \\
\ci\pi \omega\rho e^{\frac{\omega(q_2+1)}{2}} \frac{e^{-\frac{\omega\ci}{2} \sqrt{A}}
                          \left( -2 \ci + \omega \sqrt{A} \right) +
                          e^{\frac{\omega\ci}{2} \sqrt{A}}
                          \left( 2 \ci + \omega \sqrt{A} \right)}
                     {2 A^{3/2}}  & \omega < 0 \\
      0 & \omega=0 \\
\end{array}
\right.
\end{align}
\noindent with
\begin{align}\label{App:fourier:A2}
A \equiv A (\rho,z,q_2) = -(-1+q_2-2\rho-2\ci z)(-1+q_2+2\rho-2\ci z),
\end{align}
which can be simplified to a single line, given that for the Fourier transform of a real function, it holds that $F(\omega) = F^*(-\omega)$.

It is now clear from Eq. (\ref{App:fourier:Ethetadimless} - \ref{App:fourier:Hzdimless}) that the magnetic fields satisfy the necessary conditions to apply Jordan's lemma and that they have the same poles with the electric field. Hence, the exact same approach can be used leading to the following frequency domain expressions for the magnetic field
\begin{align}\label{App:fourier:Hrho_real_dimless}
H_{\rho,\mathrm{re}} (\omega) =  \left\{
\begin{array}{ll}
\left\{
\left( q_2 -1 -2 \ci z \right) \pi \rho e^{\frac{-\omega(q_2+1)}{2}}
						  \left( -\frac{e^{\frac{\omega\ci}{2} \sqrt{A}}
                          \left( -12 +6 \ci \sqrt{A} \omega + A \omega^2 \right)} {2 A^{5/2}} \right. \right. \\
                          \phantom{\left( -1 + q_2 -2 \ci z \right) \pi \rho e^{\frac{-\omega(q_2+1)}{2}}}
                          \left. \left. +
                          \frac{e^{\frac{-\omega\ci}{2} \sqrt{A}}
                          \left( -12 -6 \ci \sqrt{A} \omega + A \omega^2 \right)}
                     {2 A^{5/2}} \right)
                     \right\}^* & \omega > 0 \\
\left( q_2 -1 -2 \ci z \right) \pi \rho e^{\frac{\omega(q_2+1)}{2}}
						  \left( -\frac{e^{-\frac{\omega\ci}{2} \sqrt{A}}
                          \left( -12 -6 \ci \sqrt{A} \omega + A \omega^2 \right)} {2 A^{5/2}} \right. \\
                          \phantom{\left( -1 + q_2 -2 \ci z \right) \pi \rho e^{\frac{\omega(q_2+1)}{2}}}
                          \left. +
                          \frac{e^{\frac{\omega\ci}{2} \sqrt{A}}
                          \left( -12 +6 \ci \sqrt{A} \omega + A \omega^2 \right)}
                     {2 A^{5/2}} \right)  & \omega < 0 \\
      0 & \omega=0 \\
\end{array}
\right.
\end{align}

\begin{align}\label{App:fourier:Hz_real_dimless}
H_{z,\mathrm{re}} (\omega) =  \left\{
\begin{array}{ll}
\left\{ -\pi \ci e^{\frac{-\omega(q_2+1)}{2}} \left(
						  \frac{e^{\frac{\omega\ci}{2} \sqrt{A}}
                          \left(A r^2 \omega^2 + (A-6r^2)(2-\ci \sqrt{A} \omega)  \right)}{A^{5/2}} \right. \right. \\
                          \phantom{-\pi \ci e^{\frac{-\omega(q_2+1)}{2}}}
                          \left. \left. +
                          \frac{e^{-\frac{\omega\ci}{2} \sqrt{A}}
                          \left(-A r^2 \omega^2 + (A-6r^2)(-2-\ci \sqrt{A} \omega)  \right)}{A^{5/2}} \right)
                          \right\}^*
                           & \omega > 0 \\
-\pi \ci e^{\frac{\omega(q_2+1)}{2}} \left(
						  \frac{e^{-\frac{\omega\ci}{2} \sqrt{A}}
                          \left(A r^2 \omega^2 + (A-6r^2)(2+\ci \sqrt{A} \omega)  \right)}{A^{5/2}} \right. \\
                          \phantom{-\pi \ci e^{\frac{\omega(q_2+1)}{2}}}
                          \left. +
                          \frac{e^{\frac{\omega\ci}{2} \sqrt{A}}
                          \left(-A r^2 \omega^2 + (A-6r^2)(-2+\ci \sqrt{A} \omega)  \right)}{A^{5/2}} \right)
                           & \omega < 0 \\
      0 & \omega=0 \\
\end{array}
\right.
\end{align}

Of course one has to return to dimensional variables and insert the omitted coefficient in order to have the actual fields as follows

\begin{align}\label{App:fourier:Etheta_real}
E_{\mathrm{re}} (\omega) =  \left\{
\begin{array}{ll}
4f_0\sqrt{\mu_0/\epsilon_0}\left\{\ci\pi \frac{\omega}{c^2} \frac{\rho}{q_1} e^{-\frac{\omega(q_2+q_1)}{2c}} \frac{e^{\frac{\ci\omega}{2c} \sqrt{A}}
                          \left( 2c \ci + \omega \sqrt{A} \right) +
                          e^{-\frac{\ci\omega}{2c} \sqrt{A}}
                          \left( -2c \ci + \omega \sqrt{A} \right)}
                     {2 A^{3/2}} \right\}^* & \omega > 0 \\
4f_0\sqrt{\mu_0/\epsilon_0} \ci\pi \frac{\omega}{c^2} \frac{\rho}{q_1} e^{\frac{\omega(q_2+q_1)}{2c}} \frac{e^{-\frac{\omega\ci}{2c} \sqrt{A}}
                          \left( -2c \ci + \omega \sqrt{A} \right) +
                          e^{\frac{\omega\ci}{2c} \sqrt{A}}
                          \left( 2c \ci + \omega \sqrt{A} \right)}
                     {2 A^{3/2}}  & \omega < 0 \\
      0 & \omega=0 \\
\end{array}
\right.
\end{align}

\begin{align}\label{App:fourier:Hrho_real}
H_{\rho,\mathrm{re}} (\omega) =  \left\{
\begin{array}{ll}
\left\{
4 f_0 \pi \rho \frac{ \left(q_2 - q_1 -2 \ci z \right)}{q_1 c^2} e^{\frac{-\omega(q_2+q_1)}{2c}}
						  \left( -\frac{e^{\frac{\omega\ci}{2c} \sqrt{A}}
                          \left( -12 c^2 +6 \ci \sqrt{A} \omega c + A \omega^2 \right)} {2 A^{5/2}} \right. \right. \\
                          \phantom{4 f_0 \pi \rho \frac{ \left(q_2 - q_1 -2 \ci z \right)}{q_1 c^2} e^{\frac{-\omega(q_2+q_1)}{2c}}}
                          \left. \left. +
                          \frac{e^{\frac{-\omega\ci}{2c} \sqrt{A}}
                          \left( -12 c^2 -6 \ci \sqrt{A} \omega c + A \omega^2 \right)}
                     {2 A^{5/2}} \right)
                     \right\}^* & \omega > 0 \\
4 f_0 \pi \rho \frac{ \left(q_2 - q_1 -2 \ci z \right)}{q_1 c^2} e^{\frac{\omega(q_2+q_1)}{2c}}
						  \left( -\frac{e^{-\frac{\omega\ci}{2c} \sqrt{A}}
                          \left( -12 c^2 -6 \ci \sqrt{A} \omega c + A \omega^2 \right)} {2 A^{5/2}} \right. \\
                          \phantom{4 f_0 \pi \rho \frac{ \left(q_2 - q_1 -2 \ci z \right)}{q_1 c^2} e^{\frac{\omega(q_2+q_1)}{2c}}}
                          \left. +
                          \frac{e^{\frac{\omega\ci}{2c} \sqrt{A}}
                          \left( -12 c^2 +6 \ci \sqrt{A} \omega c + A \omega^2 \right)}
                     {2 A^{5/2}} \right)  & \omega < 0 \\
      0 & \omega=0 \\
\end{array}
\right.
\end{align}

\begin{align}\label{App:fourier:Hz_real}
H_{z,\mathrm{re}} (\omega) =  \left\{
\begin{array}{ll}
\left\{ -\frac{4\pi \ci f_0 }{q_1 c} e^{\frac{-\omega(q_2+q_1)}{2c}} \left(
						  \frac{e^{\frac{\omega\ci}{2c} \sqrt{A}}
                          \left(A r^2 \omega^2/c + (A-6r^2)(2c-\ci \sqrt{A} \omega)  \right)}{A^{5/2}}
                           \right. \right. \\
                          \phantom{-\frac{4\pi \ci f_0 }{q_1 c} e^{\frac{-\omega(q_2+q_1)}{2c}}}
                          \left. \left. +
                          \frac{e^{\frac{-\omega\ci}{2c} \sqrt{A}}
                          \left(-A r^2 \omega^2/c + (A-6r^2)(-2c-\ci \sqrt{A} \omega)  \right)}{A^{5/2}} \right)
                          \right\}^*
                           & \omega > 0 \\
-\frac{4\pi \ci f_0}{q_1 c} e^{\frac{\omega(q_2+q_1)}{2c}} \left(
						  \frac{e^{-\frac{\omega\ci}{2c} \sqrt{A}}
                          \left(A r^2 \omega^2/c + (A-6r^2)(2c+\ci \sqrt{A} \omega)  \right)}{A^{5/2}} \right. \\
                          \phantom{-\frac{4\pi \ci f_0}{q_1 c} e^{\frac{\omega(q_2+q_1)}{2c}}}
                          \left. +
                          \frac{e^{\frac{\omega\ci}{2c} \sqrt{A}}
                          \left(-A r^2 \omega^2/c + (A-6r^2)(-2c+\ci \sqrt{A} \omega)  \right)}{A^{5/2}} \right)
                           & \omega < 0 \\
      0 & \omega=0 \\
\end{array}
\right.
\end{align}
\noindent with
\begin{align}\label{App:fourier:A3}
A \equiv A (\rho,z,q_2) = -(-q_1+q_2-2\rho-2\ci z)(-q_1+q_2+2\rho-2\ci z).
\end{align}

Because now of the analyticity of the Eq. (\ref{App:fourier:Etheta}-\ref{App:fourier:Hz}), with respect to time, the real and imaginary parts (or equivalently the ``$1 \frac{1}{2}$-cycle" and the ``$1$-cycle" pulses) are Hilbert transforms of each other, which means that they share the same spectrum with a change only in phase, as it is clear from the following relations, \cite{PhysRevE.59.4630,king_2009}
\begin{align}\label{App:fourier:hilbert1}
\mathbf{E}_{\mathrm{im}} (\omega) =  \ci \, \mathrm{sgn}(\omega) \mathbf{E}_{\mathrm{re}} (\omega)
\end{align}
\noindent and
\begin{align}\label{App:fourier:hilbert2}
\mathbf{H}_{\mathrm{im}} (\omega) =  \ci \, \mathrm{sgn}(\omega) \mathbf{H}_{\mathrm{re}} (\omega),
\end{align}
\noindent with
\begin{align}\label{App:fourier:sgn}
\mathrm{sgn}(\omega) = \left\{ \begin{array}{ll}
							  1, & \omega >0 \\
							  -1, & \omega <0 \\
							  0, & \omega =0 \\
							  \end{array}
					   \right. .
\end{align}

Finally, regarding the TM pulses, the frequency domain equations can be derived by a substitution of the TE formulas to the following equations
\begin{align}
\mathbf{E}_{\mathrm{TM}} &= \sqrt{\frac{\mu_0}{\epsilon_0}} \mathbf{H}_{\mathrm{TE}} \\
\mathbf{H}_{\mathrm{TM}} &= -\sqrt{\frac{\epsilon_0}{\mu_0}} \mathbf{E}_{\mathrm{TE}}.
\end{align}

\section{Proof of isodiffraction for well-collimated doughnut pulses}
\label{App:isodiffraction}

We will now prove that in the paraxial regime of well collimated pulses, $q_2 \gg q_1$, the FD pulse can be considered as a superposition of isodiffracting beams. We will focus our study in the case of $q_2 \gg q_1$ since at $q_2 \simeq q_1$ the propagating pulse is transformed to a superposition of two counter-propagating pulses with similar energy, forming a circle of energy around the focus that is expanding towards all directions.

In order to prove that the pulse is isodiffracting we have to show that different beams diffract with the same rate. A way to do it is to show that, far from the focus, the ratio of the radial position of the maxima between any two of the monochromatic beams is independent of the propagation distance z, $\rho_{\max}(\omega_2)/\rho_{\max}(\omega_1)=\mathrm{const}$.

The spectral intensity is symmetric with respect to the $z=0$ plane which is the focal plane and hence we can prove the isodiffracting property of the pulse for $z>0$ without loss of generality. The same is true for $\omega$ and hence we will restrict our analysis to $z>0$ and $\omega >0$. It turns out that in this case the Fourier transform of the pulse is simplified to following equation
\begin{align}
E^*_{\mathrm{re}} (\omega) =
\ci\pi \omega\rho e^{-\frac{\omega(q_2+1)}{2}} \frac{
                          e^{-\frac{\ci\omega}{2} \sqrt{A}}
                          \left( -2 \ci + \omega \sqrt{A} \right)}
                     {2 A^{3/2}}.
\label{Eq:aprox_fourier}
\end{align}
That is because $\mathrm{Im} \left\{ \sqrt{A} \right\} >0$ and hence the positive exponential of Eq. \ref{Eq:fourier:real2}
\begin{align}
e^{\frac{\ci\omega}{2} \sqrt{A}} = e^{\frac{\ci\omega}{2} \mathrm{Re} \left\{ \sqrt{A} \right\}}
e^{\frac{-\omega}{2} \mathrm{Im} \left\{ \sqrt{A} \right\}}
\end{align}
is negligible compared to the negative exponential
\begin{align}
e^{\frac{-\ci\omega}{2} \sqrt{A}} = e^{-\frac{\ci\omega}{2} \mathrm{Re} \left\{ \sqrt{A} \right\}}
e^{\frac{\omega}{2} \mathrm{Im} \left\{ \sqrt{A} \right\}}
\end{align}
and thus it can be ignored.
The spectral intensity of the electric field is now given by the following equation
\begin{align}
I_{\mathrm{re}} (\omega) &= E_{\mathrm{re}} (\omega) E^*_{\mathrm{re}} (\omega)    \nonumber \\
&= \pi^2 \omega^2 \rho^2 e^{-\omega(q_2+1)}
\frac{e^{\omega \mathrm{Im} \left\{ \sqrt{A} \right\}}
\left( 4 - 4 \omega \mathrm{Im} \left\{ \sqrt{A} \right\} + \omega |A| \right)}
{2 |A|^3}.
\label{Eq:aprox_intensity}
\end{align}
For $z \gg q_2$ we can write
\begin{align}
|A| = 4(z^2+\rho^2)
\label{Eq:nA_aprox}
\end{align}
and
\begin{align}
\mathrm{Im} \left\{ \sqrt{A} \right\} &= \sqrt{|A|}\sin \left[ \frac{1}{2} \mathrm{Arg} \left( A \right) \right] \nonumber \\
&=\sqrt{|A|}\sin \left[ \frac{1}{2} \arctan \left( \frac{4(q_2-1)z}{4z^2+4\rho^2-(q_2-1)^2} \right) \right] \nonumber \\
&=\frac{(q_2-1)z}{\sqrt{z^2+\rho^2}}
\label{Eq:imrootA_aprox}
\end{align}
since $\sin(x) \simeq x$ and $\arctan (x) \simeq x$ for $x \ll 1$. In addition $|A| \gg \mathrm{Im} \left\{ \sqrt{A} \right\}$ and $|A| \gg 4$ leading to a simplified expression for the spectral intensity in the far field
\begin{align}
I_{\mathrm{re}} (\omega) &= \pi^2 \omega^3 \rho^2 e^{-\omega(q_2+1)}
\frac{e^{\frac{\omega (q_2-1)z}{\sqrt{z^2+\rho^2}}}}
{32 (z^2 + \rho^2)^2}.
\label{Eq:aprox_intensity_far_field}
\end{align}
The general shape of the spectral intensity is the same for any frequency and propagation distance, as it can be seen in Fig. \ref{Fig:fourier} e) and f), since it is given by the same equation. This equation has only one extrema which is the maximum of the intensity. That is easy to see since the function is actually the product of a decreasing exponential with the ratio of a second degree parabola and a shifted fourth degree parabola. The ratio has only one maximum and the exponential is simply shifting this maximum in space. Hence we only have to find the position of this maximum. That is the zero of the derivative of the intensity with respect to the radius, $\frac{d I_{\mathrm{re}}(\omega, \rho)}{d \rho}|_{\rho_{\max}} = 0$, which is given by the solution of the following equation
\begin{align}
4x^3-(4+\omega^2(q_2-1)^2)z^2x^2-4z^4x+4z^6=0
\label{Eq:polynomial_x}
\end{align}
with $\rho_{\max}=\sqrt{x_0}$, where $x_0$ is the only real solution of Eq. \ref{Eq:polynomial_x}.
However in the paraxial case, the radial expansion of the beam increases with a much smaller rate than the propagation distance $z$. Hence the first term of the polynomial, $x^3=\rho^6$, will be very small compared to the rest terms containing $z$ and thus it can be ignored. Finally, the equation to be solved becomes
\begin{align}
(4+\omega^2(q_2-1)^2)z^2x^2+4z^4x-4z^6=0
\label{Eq:polynomial_x_aprox}
\end{align}
and the radial position of the maximum spectral intensity in the far field is given by the following simple equation
\begin{align}
\rho_{\max} = z \sqrt{\frac{-2+2\sqrt{5+(q_2^2-1)^2\omega^2}}{4+(q_2^2-1)^2\omega^2}}.
\label{Eq:r_max}
\end{align}
The ratio of the intensity maxima of two different wavelengths can be easily obtained as
\begin{align}
\eta(\omega_2) = \frac{\rho_{\max} (\omega_2)}{\rho_{\max} (\omega_1)} = \sqrt{\frac{(4+(q_2^2-1)^2\omega_1^2)(-2+2\sqrt{5+(q_2^2-1)^2\omega_2^2})}{(-2+2\sqrt{5+(q_2^2-1)^2\omega_1^2})(4+(q_2^2-1)^2\omega_2^2)}}
\label{App:isodif:Eq:r_max_ratio}
\end{align}
and is independent of the propagation distance $z$. In addition it is easy to prove that Eq. \ref{Eq:r_max} is a monotonically decreasing function of $\omega$. The last two statements prove that the pulse is indeed described by a superposition of isodiffracting beams and that completes the proof.

\section{Hankel transform}
\label{App:hankel}

By using Jordan's lemma in a similar way with App. \ref{App:Fourier}, an analytical expression for one of the spatial coordinates of the FD pulse can be derived. However, it is clear that the pulse does not depend on the polar angle $\theta$. This symmetry can be exploited and a Hankel transform can be applied for the calculation of the spatial frequencies in a transverse plane $r, \theta$ \cite{goodman2005introduction}. However, in contrast to the scalar case, attention is required when dealing with vector fields because of the polarization. In our case the intensity of the pulse is circularly symmetric but the field is not. It has a polarization singularity at the centre leading to a sign inversion of the field across a line passing through the centre of the pulse.

In this appendix we will derive an equation for the TE pulse but a similar analysis can be performed for the TM pulse. In order to deal with the polarization we start by projecting the $\thetaunit$ dependence of field to the $\xunit, \; \yunit$ plane. The following relations between Cartesian and polar coordinates will be used

\begin{align}
\begin{array}{ll}
\rho = \sqrt{x^2 + y^2}, & \rhounit = \cos \theta \xunit + \sin \theta \yunit \\
\theta = \arctan \left( \frac{y}{x} \right), & \thetaunit = - \sin \theta \xunit + \cos \theta \yunit \\
x = \rho \cos \theta, & \xunit = \cos \theta \rhounit - \sin \theta \thetaunit \\
y = \rho \sin \theta, & \yunit = \sin \theta \rhounit + \cos \theta \thetaunit \\
\end{array}
\end{align}
for the real space and
\begin{align}
\begin{array}{ll}
k_{\rho} = \sqrt{k_x^2 + k_y^2}, & \krhounit = \cos k_{\theta} \kxunit + \sin k_{\theta} \kyunit \\
k_{\theta} = \arctan \left( \frac{k_y}{k_x} \right), & \kthetaunit = - \sin k_{\theta} \kxunit + \cos k_{\theta} \kyunit \\
k_x = k_{\rho} \cos k_{\theta}, & \kxunit = \cos k_{\theta} \krhounit - \sin k_{\theta} \kphiunit \\
k_y = k_{\rho} \sin k_{\theta}, & \kyunit = \sin k_{\theta} \krhounit + \cos k_{\theta} \kphiunit \\
\end{array}
\end{align}
for the k-space.
From the above we have
\begin{align}
{\bf E} \left(\rho, \theta \right) &= - \sin \theta  E_{\theta} \left(\rho \right) \xunit + \cos \theta  E_{\theta} \left(\rho \right) \yunit.
\label{App:hankel:Etheta_xy}
\end{align}
It is clear now that each polarization of the field is separable in the polar coordinates. In that case the Fourier transform can be expressed as an infinite sum of weighted Hankel transforms \cite{goodman2005introduction}. Let $\mathcal{F}$ and $\mathcal{H}$ denote the Fourier and Hankel transforms of a function respectively and $g(\rho, \theta) = g_{\theta} (\theta) g_{\rho} (\rho)$ being the separable function to be Fourier transformed. Then we can write
\begin{align}
\mathcal{F} \left[ g(\rho, \theta) \right] &= \sum_{-\infty}^{\infty} c_m (-\ci)^m \mathrm{e}^{\ci m k_{\theta}} \mathcal{H}_m \left[ g_{\rho} (\rho) \right],
\label{App:hankel:hankel_sum1}
\end{align}
with
\begin{align}
c_m &= \frac{1}{2 \pi} \int_0^{\infty} g_{\theta} (\theta) \mathrm{e}^{- \ci m \theta} d \theta, \label{App:hankel:hankel_sum2} \\
\mathcal{H} \left[ g_{\rho} (\rho) \right] & = 2 \pi \int_0^{\infty} \rho g_{\rho} (\rho) J_m (k_{\rho} \rho) d \rho
\label{App:hankel:hankel_sum3}
\end{align}
and $J_m$ the m order Bessel function of the first kind.
If the azimuthal part of the field $g_{\theta} (\theta)$ has some kind of azimuthal symmetry, as in our case, then only a few terms of the infinite sum will contribute to result making the problem tractable.

For the $E_x$ component we have $g_{\theta} (\theta)=-\sin \theta$ and
\begin{align}
c_m &= \frac{-1}{2 \pi} \int_0^{2 \pi} \sin \theta \mathrm{e}^{- \ci m \theta} d \theta = \frac{-1}{2 \pi} \int_0^{2 \pi} \frac{\mathrm{e}^{- \ci \theta (m-1)}}{2 \ci} d \theta + \frac{1}{2 \pi} \int_0^{2 \pi} \frac{\mathrm{e}^{- \ci \theta (m+1)}}{2 \ci} d \theta \nonumber \\
&= \left\{
\begin{array}{ll}
-\frac{1}{2 \ci}, & m=1 \\
\frac{1}{2 \ci}, & m=-1 \\
0, & m\neq \pm 1 \\
\end{array}
\right.
\label{App:hankel:cmx}
\end{align}
and for the $E_y$, $g_{\theta} (\theta)= \cos \theta$ and
\begin{align}
c_m &= \left\{
\begin{array}{ll}
\frac{1}{2} , & m \pm 1 \\
0, & m\neq \pm 1 \\
\end{array}
\right. .
\label{App:hankel:cmy}
\end{align}
Hence, from Eq. (\ref{App:hankel:Etheta_xy} - \ref{App:hankel:cmy}) and by using the identity $J_{-n} (x) = (-1)^n J_n (x)$ we have
\begin{align}
E_x \left(k_{\rho}, k_{\theta} \right) \xunit &= 2 \pi \ci \sin k_{\theta} \int_0^{\infty} \rho E_{\theta} \left(\rho \right) J_1 (k_{\rho} \rho) d \rho \xunit
\end{align}
and
\begin{align}
E_y \left(k_{\rho}, k_{\theta} \right) \yunit &= -2 \pi \ci \cos k_{\theta} \int_0^{\infty} \rho E_{\theta} \left(\rho \right) J_1 (k_{\rho} \rho) d \rho \yunit.
\end{align}
However, since $\xunit,\yunit,\thetaunit,\kxunit,\kyunit$ and $\kthetaunit$ are unit vectors in the same coordinate system, $\xunit = \kxunit$, $\yunit = \kyunit$ and $\thetaunit = \kthetaunit$ leading to a single azimuthally polarized equation without a $k_{\theta}$ dependence
\begin{align}
{\bf E} \left(k_{\rho} \right) &= -2 \pi \ci \int_0^{\infty} \rho E_{\theta} \left(\rho \right) J_1 (k_{\rho} \rho) d \rho \thetaunit.
\label{App:hankel:forward_hankel}
\end{align}

For the inverse transform we have to look first at the definition of the forward transform that was used to derive Eq. \ref{App:hankel:forward_hankel},
\begin{align}
{\bf E} \left(k_{\rho}, k_{\theta} \right) &= \int_0^{2 \pi} \int_0^\infty - \sin \theta \mathrm{e}^{-i k_{\rho} \rho (\cos \theta \cos k_{\theta} + \sin \theta \sin k_{\theta}) } E_{\theta} \left(\rho \right) \rho d \rho d \theta \xunit \nonumber \\
&+ \int_0^{2 \pi} \int_0^\infty \cos \theta \mathrm{e}^{-i k_{\rho} \rho (\cos \theta \cos k_{\theta} + \sin \theta \sin k_{\theta}) } E_{\theta} \left(\rho \right) \rho d \rho d \theta \yunit.
\label{App:hankel:inv_def1}
\end{align}
In the same manner we can write the inverse as
\begin{align}
{\bf E} \left( \rho, \theta \right) &= \frac{1}{4 \pi^2} \int_0^{2 \pi} \int_0^\infty - \sin k_{\theta} \mathrm{e}^{i k_{\rho} \rho (\cos \theta \cos k_{\theta} + \sin \theta \sin k_{\theta}) } E_{\theta} \left(k_{\rho} \right) k_{\rho} d k_{\rho} d k_{\theta} \kxunit \nonumber \\
&+ \frac{1}{4 \pi^2} \int_0^{2 \pi} \int_0^\infty \cos k_{\theta} \mathrm{e}^{i k_{\rho} \rho (\cos \theta \cos k_{\theta} + \sin \theta \sin k_{\theta}) } E_{\theta} \left(k_{\rho} \right) k_{\rho} d k_{\rho} d k_{\theta} \kyunit.
\label{App:hankel:inv_def2}
\end{align}
Finally, we define $\theta' = \theta - \pi$, hence $\sin (\theta' + \pi) = -\sin \theta'$ and $\cos (\theta' + \pi) = -\cos \theta'$. The integral limits change to $(- \pi, \pi)$ but since the integrand has a period of $2 \pi$ this does not change the integral and hence there is no need to change the limits. By ignoring the prime at $\theta'$ for clarity we end up with the following equation for the inverse Fourier transform
\begin{align}
{\bf E} \left( \rho, \theta \right) &= \frac{1}{4 \pi^2} \int_0^{2 \pi} \int_0^\infty  \sin k_{\theta} \mathrm{e}^{-i k_{\rho} \rho (\cos \theta \cos k_{\theta} + \sin \theta \sin k_{\theta}) } E_{\theta} \left(k_{\rho} \right) k_{\rho} d k_{\rho} d k_{\theta} \kxunit \nonumber \\
&+ \frac{1}{4 \pi^2} \int_0^{2 \pi} \int_0^\infty - \cos k_{\theta} \mathrm{e}^{-i k_{\rho} \rho (\cos \theta \cos k_{\theta} + \sin \theta \sin k_{\theta}) } E_{\theta} \left(k_{\rho} \right) k_{\rho} d k_{\rho} d k_{\theta} \kyunit.
\label{App:hankel:inv_def3}
\end{align}
It is clear now that by swapping $\rho$ with $k_{\rho}$ and $\theta$ with $k_{\theta}$, Eq. \ref{App:hankel:inv_def3} is identical to Eq. \ref{App:hankel:inv_def1} except of the coefficient $\frac{1}{4 \pi^2}$ and a minus sign. Hence the inverse Fourier transform can be given from the following Hankel transform
\begin{align}
{\bf E} \left(\rho \right) &=  \frac{\ci}{2 \pi} \int_0^{\infty} k_{\rho} E_{\theta} \left(k_{\rho} \right) J_1 (k_{\rho} \rho) d k_{\rho} \thetaunit.
\label{App:hankel:inverse_hankel}
\end{align}

Returning to Eq. \ref{App:hankel:forward_hankel}, we will solve the integral for the complex expression of the field. The transforms of the real and imaginary fields are then simply taken from the following equations

\begin{align}
E_{\mathrm{re}, \theta} (k_{\rho}) &= \frac{ E_{\theta} (k_{\rho}) -  E^*_{\theta} (k_{\rho})}{2} \\
E_{\mathrm{im}, \theta} (k_{\rho}) &= \frac{ E_{\theta} (k_{\rho}) +  E^*_{\theta} (k_{\rho})}{2 \ci}.
\end{align}

\begin{proof}
\begin{align*}
E_{\mathrm{re}, \theta} (k_{\rho}) &= -2 \pi \ci \int_0^{\infty} \mathrm{Re} \left\{ E_{\theta} (\rho) \right\} J_1(k_{\rho} \rho)\rho d \rho \nonumber \\
&=  \int_0^{\infty}  \frac{-2 \pi \ci E_{\theta} (\rho)}{2} J_1(k_{\rho} \rho)\rho d \rho - \left[ \int_0^{\infty}  \frac{-2 \pi \ci E_{\theta} (\rho)}{2}  J_1(k_{\rho} \rho)\rho d \rho \right]^* \nonumber \\
&= \frac{ E_{\theta} (k_{\rho}) -  E^*_{\theta} (k_{\rho})}{2}
\end{align*}
and similar for the imaginary part, given that $k_{\rho}$ and $\rho$ are real.
\end{proof}

For clarity we rewrite the electric field expression as follows
\begin{align}
E_{\theta} &=-4 \ci f_0 \sqrt{\frac{\mu_0}{\epsilon_0}} \frac{\rho \left( q_1 +q_2 -2\ci c t \right)}{\left[ \rho^2 + (q_1 +\ci z -\ci ct)(q_2 -\ci z - \ci ct) \right]^3}  \nonumber \\
&= \mathrm{coef} \frac{\rho}{\left( \rho^2 + \alpha^2 \right)^3}
\label{App:hankel:Etheta_init}
\end{align}
\noindent with $coef = -4 \ci f_0 \sqrt{\frac{\mu_0}{\epsilon_0}} (q_1 + q_2 -2 \ci ct)$ and $\alpha = \sqrt{(q_1 +\ci z -\ci ct)(q_2 -\ci z - \ci ct)}$. The integral that we have to solve now becomes

\begin{align}
E_{\theta} (k_{\rho}) &= - 2 \pi \ci coef \int_0^{\infty} \frac{\rho^2}{\left( \rho^2 + \alpha^2 \right)^3} J_1(k_{\rho} \rho) d \rho.
\label{App:hankel:hank_Etheta2}
\end{align}

The solution of an integral of the above form is known when it satisfies some criteria and it is given from the following form \cite{Lin2014, watson1944treatise}
\begin{align}
\int_0^{\infty} \frac{\rho^{\nu+1}}{\left( \rho^2 + \alpha^2 \right)^{\mu+1}} J_{\nu}(k \rho) d \rho = \frac{k^{\mu} \alpha^{\nu - \mu}}{2^{\mu} \Gamma(\mu+1)} K_{\nu-\mu} (k \alpha)
\label{App:hankel:hank_integral}
\end{align}
with $\mathrm{Re} \left\{ \alpha \right\} > 0$ and $-1< \mathrm{Re} \left\{ \nu \right\} < 2 \mathrm{Re} \left\{ \mu \right\} + 3/2$. $K_{\nu}$ is the $\nu$ order modified Bessel function of the second kind and $\Gamma$ denotes the Gamma function with its integral definition being \cite{arfken2012mathematical}
\begin{align}
\Gamma (z) = \int_0^{\infty} t^{z-1} e^t dt, \; \mathrm{Re} \left\{ z \right\} >0.
\end{align}
The only properties of this function that we will need here are
\begin{align}
\Gamma (z+1) &= z \Gamma (z), \\
\Gamma (1) &= 1.
\end{align}
In our case, $\nu = 1$ and $\mu=2$ and hence the second criterion is satisfied. The first criterion requires $\mathrm{Re} \left\{\alpha \right\} > 0$ or $|\mathrm{\arg} (\alpha)| < \pi/2$. We will now show that $|\mathrm{\arg} (\alpha)| < \pi/2$ is true in our case and hence all the conditions needed to apply the above formula are met.

\begin{proof}
We have,
\begin{align*}
\alpha &= \sqrt{q_1 q_2 +z^2-c^2t^2-\ci \left( ct(q_1+q_2) + z(q_1-q_2) \right)} \\
\beta &= \alpha^2 = q_1 q_2 +z^2-c^2t^2-\ci \left( ct(q_1+q_2) + z(q_1-q_2) \right).
\end{align*}
We want to study when $| \mathrm{arg} (\alpha) | < \pi/2$ or, since $\alpha= \sqrt{\beta} = \sqrt{|\beta|e^{\ci \theta}} = \sqrt{|\beta|}e^{\ci \theta/2}$, when  $| \mathrm{arg} (\beta) | = | \mathrm{arg} (\alpha^2) | < \pi$. We write $\beta = x+y \ci$ and hence $\mathrm{arg} (\beta) = \mathrm{atan2}(y,x)$. From Eq. \ref{App:fourier:atan2} there is only one case that the criterion can be possibly violated, the case $\mathrm{arctan}(y/x)+\pi$ with $x<0, y \geq 0$. But $-\pi/2<\mathrm{arctan}(y/x) \leq 0$ and hence the equality $| \mathrm{arg} (\alpha) | = \pi/2$ can only happen when $y=0$ and $x<0$.

For $y=0$ we have
\begin{align}
ct = \frac{q_2-q_1}{q_2+q_1}z.
\label{App:hankel:cond1}
\end{align}

For $x<0$ we have
\begin{align}
c^2t^2 > q_1q_2 +z^2.
\label{App:hankel:cond2}
\end{align}
Substitution of Eq. \ref{App:hankel:cond1} to \ref{App:hankel:cond2} leads to the following statement
\begin{align}
\frac{-4q_1q_2}{(q_1+q_2)^2}z^2 > q_1q_2
\end{align}
which is not true since $q_1$ and $q_2$ are both positive numbers. Hence,  $| \mathrm{arg} (\alpha) |$ is always smaller than $\pi/2$.
\end{proof}

The Hankel transform is now given by the equation
\begin{align}
E_{\theta} (k_{\rho}) &= - 2 \pi \ci coef \frac{k^2_{\rho}}{8a} K_{-1} (k_{\rho} \alpha).
\label{App:hankel:hank_Etheta3}
\end{align}
From the definition of the $K_{\nu}$ it can be easily shown that $K_{-1}(x) = K_1(x)$, hence we finally have
\begin{align}
\mathbf{E} (k_{\rho}) &= - 8 \pi f_0 \sqrt{\frac{\mu_0}{\epsilon_0}} (q_1 + q_2 -2 \ci ct) \frac{k^2_{\rho}}{8a} K_{1} (k_{\rho} \alpha) \thetaunit,
\label{App:hankel:hank_Etheta4}
\end{align}
with $\alpha = \sqrt{(q_1 +\ci z -\ci ct)(q_2 -\ci z - \ci ct)}$.

\section{Demposition of FD pulses into single-cycle pulses}
\label{App:slice}

\setcounter{subsection}{0}
\renewcommand{\thesubsection}{\Alph{subsection}}

\subsection{Luminal Filter in Spacetime domain\label{app:SpaceTimeFilter}}

We want a filter that would remove all components that do not match
the wave equation, i.e. we want to keep only $\omega/c=k$ components.
Let's assume such filter exists and is denoted by $\tilde{h}\left(\omega,\,\vec{k}\right)$
in frequency domain. We want to get the expression for it in spacetime
domain:

\[
h\left(t,\,\vec{r}\right)=\frac{1}{2\pi}\int d\omega\,\exp\left(-i\omega t\right)\,\left(\frac{1}{2\pi}\right)^{3}\int d^{3}k\,\exp\left(i\vec{k}.\vec{r}\right)\tilde{h}
\]

Firstly, we expect the filter to be independent of the direction of
the wavevector in frequency space, and of direction of position in
real space, so:
\begin{flalign*}
h\left(t,\,\vec{r}\right)=h\left(t,\,r\right)= & \frac{1}{2\pi}\int d\omega\,\exp\left(-i\omega t\right)\,\left(\frac{1}{2\pi}\right)^{3}\int_{0}^{\infty}dk\,k^{2}\tilde{h}\left(\omega,\,\left|\vec{k}\right|\right)\cdot\\
&\quad\quad\cdot\int_{0}^{\pi}d\theta\,\sin\theta\exp\left(ikr\cos\theta\right)\int_{0}^{2\pi}d\phi\,1\\
= & \frac{1}{2\pi r}\int d\omega\,\exp\left(-i\omega t\right)\,\frac{1}{2\pi^{2}}\int_{0}^{\infty}dk\,\tilde{h}\left(\omega,\,\left|\vec{k}\right|\right)\,k\,\sin\left(kr\right)\\
= & \frac{1}{2\pi r}\int d\omega\,\exp\left(-i\omega t\right)\,\frac{1}{4\pi^{2}}\int_{-\infty}^{\infty}dk\,\tilde{h}\left(\omega,\,\vec{k}\right)\,k\,\sin\left(kr\right)
\end{flalign*}

In the last step we extended the integral to span negative $k$, since
it is convenient and since the integrand does not change with this
sign of $k$. We can now apply the limitation:
\[
\tilde{h}\left(\dots\right)=\lim_{\epsilon\to0}\begin{cases}
\alpha, & -\frac{\epsilon}{2}<\left(\left|k\right|-\frac{\omega}{c}\right)<\frac{\epsilon}{2}\\
\alpha, & -\frac{\epsilon}{2}<\left(\left|k\right|+\frac{\omega}{c}\right)<\frac{\epsilon}{2}\\
0, & otherwise
\end{cases}
\]

Where we have left freedom to adjust the constant $\alpha$ for normalization
purposes. Thus:
\begin{flalign*}
h\left(t,\,r\right)= & \frac{1}{2\pi r}\int d\omega\,\exp\left(-i\omega t\right)\,\frac{1}{4\pi^{2}}\int_{-\infty}^{\infty}dk\,\tilde{h}\left(\omega,\,\left|\vec{k}\right|\right)\,k\,\sin\left(kr\right)\\
= & \frac{1}{8\pi^{3}r}\int d\omega\,\exp\left(-i\omega t\right)\,\lim_{\epsilon\to0}\alpha\left(\int_{\frac{\omega}{c}-\frac{\epsilon}{2}}^{\frac{\omega}{c}+\frac{\epsilon}{2}}dk\,k\,\sin\left(kr\right)+\int_{-\frac{\omega}{c}-\frac{\epsilon}{2}}^{-\frac{\omega}{c}+\frac{\epsilon}{2}}dk\,k\,\sin\left(kr\right)\right)\\
= & \frac{1}{4\pi^{3}r}\lim_{\epsilon\to0}\alpha\int d\omega\int_{\frac{\omega}{c}-\frac{\epsilon}{2}}^{\frac{\omega}{c}+\frac{\epsilon}{2}}dk\,\left(k\,\sin\left(kr\right)\exp\left(-i\omega t\right)\right)
\end{flalign*}

This essentially is an area integral in the $\omega k$-space. The
actual domain of integration is a small space around the $\omega=ck$
line. Let us reparameterize this:
\begin{flalign*}
k= & \omega/c+\zeta
\end{flalign*}

Then:
\begin{flalign*}
h\left(t,\,r\right)= & \frac{1}{4\pi^{3}r}\lim_{\epsilon\to0}\alpha\int d\omega\int_{-\frac{\epsilon}{2}}^{\frac{\epsilon}{2}}d\zeta\,\left(\left(\frac{\omega}{c}+\zeta\right)\,\sin\left(\left(\frac{\omega}{c}+\zeta\right)r\right)\exp\left(-i\omega t\right)\right)\\
= & \frac{1}{4\pi^{3}r}\lim_{\epsilon\to0}\int_{-\frac{\epsilon}{2}}^{\frac{\epsilon}{2}}d\zeta\frac{1}{i2}\int d\omega\:\alpha\left(\frac{\omega}{c}+\zeta\right)\exp\left(-i\omega t\right)\cdot\\
&\quad\quad\cdot\left(\exp\left(i\left(\frac{\omega}{c}+\zeta\right)r\right)-\exp\left(-i\left(\frac{\omega}{c}+\zeta\right)r\right)\right)\\
= & \frac{1}{i8\pi^{3}r}\lim_{\epsilon\to0}\int_{-\frac{\epsilon}{2}}^{\frac{\epsilon}{2}}d\zeta\int d\omega\:\alpha\left(\frac{\omega}{c}+\zeta\right)\cdot\\
&\quad\quad\cdot\left(\exp\left(i\omega\left(\frac{r}{c}-t\right)\right)\exp\left(i\zeta r\right)-\exp\left(-i\omega\left(\frac{r}{c}+t\right)\right)\exp\left(-i\zeta r\right)\right)\\
= & \frac{1}{i8\pi^{3}r}\lim_{\epsilon\to0}\int_{-\frac{\epsilon}{2}}^{\frac{\epsilon}{2}}d\zeta\alpha\left(\frac{\partial_{t}}{-ic}+\zeta\right)\cdot\\
&\quad\quad\cdot\int d\omega\left(\exp\left(i\omega\left(\frac{r}{c}-t\right)\right)\exp\left(i\zeta r\right)-\exp\left(-i\omega\left(\frac{r}{c}+t\right)\right)\exp\left(-i\zeta r\right)\right)
\end{flalign*}

Now we can evaluate the integrals to get the delta functions:

\begin{flalign*}
h\left(t,\,r\right)= & \frac{1}{i8\pi^{3}r}\lim_{\epsilon\to0}\int_{-\frac{\epsilon}{2}}^{\frac{\epsilon}{2}}d\zeta\alpha\left(\frac{\partial_{t}}{-ic}+\zeta\right)\cdot\\
&\quad\quad\cdot\int d\omega\left(\exp\left(i\omega\left(\frac{r}{c}-t\right)\right)\exp\left(i\zeta r\right)-\exp\left(-i\omega\left(\frac{r}{c}+t\right)\right)\exp\left(-i\zeta r\right)\right)\\
= & \frac{1}{i8\pi^{3}r}\lim_{\epsilon\to0}\int_{-\frac{\epsilon}{2}}^{\frac{\epsilon}{2}}d\zeta\alpha\left(\frac{\partial_{t}}{-ic}+\zeta\right)\left(2\pi\delta\left(\frac{r}{c}-t\right)\exp\left(i\zeta r\right)-2\pi\delta\left(\frac{r}{c}+t\right)\exp\left(-i\zeta r\right)\right)\\
= & \frac{1}{4\pi^{2}rc}\lim_{\epsilon\to0}\int_{-\frac{\epsilon}{2}}^{\frac{\epsilon}{2}}d\zeta\alpha\left(\partial_{t}-ic\zeta\right)\left(\delta\left(\frac{r}{c}-t\right)\exp\left(i\zeta r\right)-\delta\left(\frac{r}{c}+t\right)\exp\left(-i\zeta r\right)\right)
\end{flalign*}

Thus, up to constant $\alpha$, the filter is:
\begin{multline}
h\left(t-t',\,\left|\vec{r}-\vec{r}'\right|\right)=\frac{1}{4\pi^{2}c}\lim_{\epsilon\to0}\alpha\int_{-\frac{\epsilon}{2}}^{\frac{\epsilon}{2}}d\zeta\left(\partial_{t}-ic\zeta\right)\cdot\\
\quad\quad\cdot\frac{\delta\left(\frac{\left|\vec{r}-\vec{r}'\right|}{c}-t+t'\right)\exp\left(i\zeta\left|\vec{r}-\vec{r}'\right|\right)-\delta\left(\frac{\left|\vec{r}-\vec{r}'\right|}{c}+t-t'\right)\exp\left(-i\zeta\left|\vec{r}-\vec{r}'\right|\right)}{\left|\vec{r}-\vec{r}'\right|}\label{eq:filterPreDef}
\end{multline}

\subsubsection{Fix normalization: filter the plane wave}

The only requirement we have is that:
\begin{equation}
h\otimes\exp\left(i\left(\omega t-\vec{k}.\vec{r}\right)\right)=\begin{cases}
\exp\left(i\left(\omega t-\vec{k}.\vec{r}\right)\right), & \omega=\pm c\sqrt{\vec{k}.\vec{k}}\:\&\:k\neq0\\
??? & \omega=0\,\&\,k=0\\
0, & otherwise
\end{cases}\label{eq:FilterReq}
\end{equation}

Lets check this:

\begin{flalign}
&h\otimes\exp\left(i\left(\omega t-\vec{k}.\vec{r}\right)\right)=  \frac{1}{4\pi^{2}c}\lim_{\epsilon\to0}\alpha\int_{-\frac{\epsilon}{2}}^{\frac{\epsilon}{2}}d\zeta\left(\partial_{t}-ic\zeta\right)\int dt'\exp\left(i\omega t'\right)\int d^{3}r'\,\exp\left(-i\vec{k}.\vec{r}'\right)\cdot\nonumber \\
 &\quad\quad\quad \cdot\frac{\delta\left(\frac{\left|\vec{r}-\vec{r}'\right|}{c}-t+t'\right)\exp\left(i\zeta\left|\vec{r}-\vec{r}'\right|\right)-\delta\left(\frac{\left|\vec{r}-\vec{r}'\right|}{c}+t-t'\right)\exp\left(-i\zeta\left|\vec{r}-\vec{r}'\right|\right)}{\left|\vec{r}-\vec{r}'\right|}\label{eq:FullFilterEq}
\end{flalign}

We will start with the most common case, and then analyze two special cases.

\paragraph{For $\omega\protect\neq0$ and $\vec{k}\protect\neq\vec{0}$}

Starting from Eq.~(\ref{eq:FullFilterEq}) and applying the time-integral:
\begin{flalign*}
h\otimes\exp\left(i\left(\omega t-\vec{k}.\vec{r}\right)\right)= & \frac{-i}{4\pi^{2}}\lim_{\epsilon\to0}\alpha\int_{-\frac{\epsilon}{2}}^{\frac{\epsilon}{2}}d\zeta\left(\zeta-\frac{\partial_{t}}{ic}\right)\exp\left(i\omega t\right)\int d^{3}r'\,\exp\left(-i\vec{k}.\vec{r}'\right)\cdot\\
 & \cdot\frac{\exp\left(i\left(\zeta-\frac{\omega}{c}\right)\left|\vec{r}-\vec{r}'\right|\right)-\delta\left(\frac{\left|\vec{r}-\vec{r}'\right|}{c}+t-t'\right)\exp\left(-i\left(\zeta-\frac{\omega}{c}\right)\left|\vec{r}-\vec{r}'\right|\right)}{\left|\vec{r}-\vec{r}'\right|}\\
= & \frac{-i\exp\left(i\omega t\right)}{4\pi^{2}}\lim_{\epsilon\to0}\alpha\int_{-\frac{\epsilon}{2}}^{\frac{\epsilon}{2}}d\zeta\left(\zeta-\frac{\omega}{c}\right)\int d^{3}r'\,\exp\left(-i\vec{k}.\vec{r}'\right)\cdot\\
 & \cdot i2\frac{\sin\left(\sigma\left|\vec{r}-\vec{r}'\right|\right)}{\left|\vec{r}-\vec{r}'\right|}\\
= & \frac{\exp\left(i\omega t\right)}{2\pi^{2}}\lim_{\epsilon\to0}\alpha\int_{-\frac{\epsilon}{2}-\frac{\omega}{c}}^{\frac{\epsilon}{2}-\frac{\omega}{c}}d\sigma\,\sigma\int d^{3}r'\,\exp\left(-i\vec{k}.\vec{r}'\right)\cdot\frac{\sin\left(\sigma\left|\vec{r}-\vec{r}'\right|\right)}{\left|\vec{r}-\vec{r}'\right|}\\
= & \frac{\exp\left(i\omega t\right)}{2\pi^{2}}\lim_{\epsilon\to0}\alpha\int_{-\frac{\epsilon}{2}-\frac{\omega}{c}}^{\frac{\epsilon}{2}-\frac{\omega}{c}}d\sigma\,\sigma I_{-}\left(\sigma,\,\vec{r},\,\vec{k}\right)
\end{flalign*}

Where we used:
\[
I_{\pm}\left(\sigma,\,\vec{r},\,\vec{k}\neq\vec{0}\right)=\int d^{3}r'\,\exp\left(\pm i\vec{k}.\vec{r}'\right)\cdot\frac{\sin\left(\sigma\left|\vec{r}-\vec{r}'\right|\right)}{\left|\vec{r}-\vec{r}'\right|}
\]

Which is treated in App.~\ref{app:Iint}. One can see why distinction
$\vec{k}\neq\vec{0}$ is important. Since the integration limit goes
to infinity, and since integrand does not converge, no matter how
small is the magnitude of $\vec{k}$ it can still make an important
contribution. The case $\vec{k}=0$ has to be treated independently.
In case of $\omega\neq0$ the integration variable $\sigma\neq0$,
so using Eq.~(\ref{eq:FullIInt}):

\begin{flalign*}
h\otimes\exp\left(i\left(\omega t-\vec{k}.\vec{r}\right)\right)= & \frac{\exp\left(i\omega t\right)}{2\pi^{2}}\lim_{\epsilon\to0}\alpha\int_{-\frac{\epsilon}{2}-\frac{\omega}{c}}^{\frac{\epsilon}{2}-\frac{\omega}{c}}d\sigma\,\sigma\cdot\\
&\quad\quad\cdot\left(\frac{2\pi^{2}}{\sigma}\left(\delta\left(\sigma-k\right)+\delta\left(\sigma+k\right)\right)\cdot\exp\left(-i\vec{k}.\vec{r}\right)\right)\\
= & \exp\left(i\left(\omega t-\vec{k}.\vec{r}\right)\right)\lim_{\epsilon\to0}\alpha\int_{-\frac{\epsilon}{2}-\frac{\omega}{c}}^{\frac{\epsilon}{2}-\frac{\omega}{c}}d\sigma\,\left(\delta\left(\sigma-k\right)+\delta\left(\sigma+k\right)\right)\\
= & \begin{cases}
\alpha\exp\left(i\left(\omega t-\vec{k}.\vec{r}\right)\right), & \omega=\pm ck,\,\omega\neq0,\,k\neq0\\
0, & otherwise
\end{cases}
\end{flalign*}

So this works as long as $\alpha=1$. Now the special cases.

\paragraph{Check for $\omega=0$}

We start from:

\begin{flalign*}
h\otimes\exp&\left(i\left(-\vec{k}.\vec{r}\right)\right)= \\
= & \frac{1}{4\pi^{2}c}\lim_{\epsilon\to0}\alpha\int_{-\frac{\epsilon}{2}}^{\frac{\epsilon}{2}}d\zeta\left(\partial_{t}-ic\zeta\right)\int dt'\int d^{3}r'\,\exp\left(-i\vec{k}.\vec{r}'\right)\cdot\\
 & \cdot\frac{\delta\left(\frac{\left|\vec{r}-\vec{r}'\right|}{c}-t+t'\right)\exp\left(i\zeta\left|\vec{r}-\vec{r}'\right|\right)-\delta\left(\frac{\left|\vec{r}-\vec{r}'\right|}{c}+t-t'\right)\exp\left(-i\zeta\left|\vec{r}-\vec{r}'\right|\right)}{\left|\vec{r}-\vec{r}'\right|}\\
= & \frac{-i}{4\pi^{2}}\lim_{\epsilon\to0}\alpha\int_{-\frac{\epsilon}{2}}^{\frac{\epsilon}{2}}d\zeta\zeta\int d^{3}r'\,\exp\left(-i\vec{k}.\vec{r}'\right)\cdot\frac{\exp\left(i\zeta\left|\vec{r}-\vec{r}'\right|\right)-\exp\left(-i\zeta\left|\vec{r}-\vec{r}'\right|\right)}{\left|\vec{r}-\vec{r}'\right|}\\
= & \frac{-i\cdot i2}{4\pi^{2}}\lim_{\epsilon\to0}\alpha\int_{-\frac{\epsilon}{2}}^{\frac{\epsilon}{2}}d\zeta\zeta\int d^{3}r'\,\exp\left(-i\vec{k}.\vec{r}'\right)\cdot\frac{\sin\left(\zeta\left|\vec{r}-\vec{r}'\right|\right)}{\left|\vec{r}-\vec{r}'\right|}\\
= & \frac{1}{2\pi^{2}}\lim_{\epsilon\to0}\alpha\int_{-\frac{\epsilon}{2}}^{\frac{\epsilon}{2}}d\zeta\zeta I_{-}\left(\zeta,\,\vec{r},\,\vec{k}\right)
\end{flalign*}

Now we shall first use the Eq.~(\ref{eq:FullIInt}) and then take
the limit $k\to0$, the meaning of this is simple: $k$ is not zero
for purposes of $I_{-}\left(\zeta,\,\vec{r},\,\vec{k}\right)$ but
it is within the range $ck=\pm\epsilon/2$, but since $\epsilon\to0$
is to be the last limit, we can proceed:
\begin{flalign*}
h\otimes\exp\left(-i\vec{k}.\vec{r}\right)= & \frac{1}{2\pi^{2}}\lim_{\epsilon\to0}\alpha\left(\text{PV}\int_{-\frac{\epsilon}{2}}^{\frac{\epsilon}{2}}d\zeta\zeta\frac{2\pi^{2}}{\zeta}\left(\delta\left(\zeta-k\right)+\delta\left(\zeta+k\right)\right)\cdot\exp\left(-i\vec{k}.\vec{r}\right)\right)
\end{flalign*}

Here we took care of the change in the value of $I_{-}$ when $\sigma=0$
by setting one integral to be `principal value'. Moving on:

\begin{flalign*}
h\otimes\exp\left(-i\vec{k}.\vec{r}\right)= & \lim_{\epsilon\to0}\alpha\text{PV}\int_{-\frac{\epsilon}{2}}^{\frac{\epsilon}{2}}d\zeta\left(\delta\left(\zeta-k\right)+\delta\left(\zeta+k\right)\right)\\
= & \alpha\lim_{\epsilon\to0^{+}}\begin{cases}
2, & 0<k<\epsilon/2\\
0, & otherwise
\end{cases}=\alpha\begin{cases}
1 & k\to0^{+}\\
0, & otherwise
\end{cases}
\end{flalign*}

\paragraph{Check for $\vec{k}=0$}

Start from:
\begin{flalign*}
h\otimes\exp\left(i\omega t\right)= & \frac{1}{4\pi^{2}c}\lim_{\epsilon\to0}\alpha\int_{-\frac{\epsilon}{2}}^{\frac{\epsilon}{2}}d\zeta\left(\partial_{t}-ic\zeta\right)\int dt'\exp\left(i\omega t'\right)\int d^{3}r'\,\cdot\\
 & \cdot\frac{\delta\left(\frac{\left|\vec{r}-\vec{r}'\right|}{c}-t+t'\right)\exp\left(i\zeta\left|\vec{r}-\vec{r}'\right|\right)-\delta\left(\frac{\left|\vec{r}-\vec{r}'\right|}{c}+t-t'\right)\exp\left(-i\zeta\left|\vec{r}-\vec{r}'\right|\right)}{\left|\vec{r}-\vec{r}'\right|}\\
= & \frac{1}{4\pi^{2}c}\lim_{\epsilon\to0}\alpha\int_{-\frac{\epsilon}{2}}^{\frac{\epsilon}{2}}d\zeta\left(\partial_{t}-ic\zeta\right)\exp\left(i\omega t\right)\int d^{3}r'\,\cdot\\
 & \cdot\frac{\exp\left(i\left(\zeta-\frac{\omega}{c}\right)\left|\vec{r}-\vec{r}'\right|\right)-\exp\left(-i\left(\zeta-\frac{\omega}{c}\right)\left|\vec{r}-\vec{r}'\right|\right)}{\left|\vec{r}-\vec{r}'\right|}\\
= & \frac{-i\exp\left(i\omega t\right)}{4\pi^{2}}\lim_{\epsilon\to0}\alpha\int_{-\frac{\epsilon}{2}}^{\frac{\epsilon}{2}}d\zeta\left(\zeta-\frac{\omega}{c}\right)\int d^{3}r'\frac{i2\sin\left(\left(\zeta-\frac{\omega}{c}\right)\left|\vec{r}-\vec{r}'\right|\right)}{\left|\vec{r}-\vec{r}'\right|}\\
= & \frac{\exp\left(i\omega t\right)}{2\pi^{2}}\lim_{\epsilon\to0}\alpha\int_{-\frac{\epsilon}{2}-\frac{\omega}{c}}^{\frac{\epsilon}{2}-\frac{\omega}{c}}d\sigma\sigma\int d^{3}r'\frac{\sin\left(\sigma\left|\vec{r}-\vec{r}'\right|\right)}{\left|\vec{r}-\vec{r}'\right|}\\
= & \frac{\exp\left(i\omega t\right)}{2\pi^{2}}\lim_{\epsilon\to0}\alpha\int_{-\frac{\epsilon}{2}-\frac{\omega}{c}}^{\frac{\epsilon}{2}-\frac{\omega}{c}}d\sigma\sigma I\left(\sigma,\,\vec{r}\right)
\end{flalign*}

Where:
\[
I\left(\sigma,\,\vec{r}\right)=\int d^{3}r'\frac{\sin\left(\sigma\left|\vec{r}-\vec{r}'\right|\right)}{\left|\vec{r}-\vec{r}'\right|}
\]

Using Eq.~(\ref{eq:IintK0}):
\begin{flalign*}
h\otimes\exp\left(i\omega t\right)= & \frac{\exp\left(i\omega t\right)}{2\pi^{2}}\lim_{\epsilon\to0}\alpha\int_{-\frac{\epsilon}{2}-\frac{\omega}{c}}^{\frac{\epsilon}{2}-\frac{\omega}{c}}d\sigma\sigma\left(-4\pi^{2}\frac{\sin\left(\sigma r\right)}{r\sigma}\delta'\left(\sigma\right)\right)\\
= & -\frac{2}{r}\exp\left(i\omega t\right)\lim_{\epsilon\to0}\alpha\int_{-\frac{\epsilon}{2}-\frac{\omega}{c}}^{\frac{\epsilon}{2}-\frac{\omega}{c}}d\sigma\sin\left(\sigma r\right)\delta'\left(\sigma\right)\\
= & -\frac{2\alpha}{r}\exp\left(i\omega t\right)\lim_{\epsilon\to0}\lim_{\eta\to0}\frac{1}{\eta}\int_{-\frac{\epsilon}{2}-\frac{\omega}{c}}^{\frac{\epsilon}{2}-\frac{\omega}{c}}d\sigma\sin\left(\sigma r\right)\left(\delta\left(\sigma+\eta/2\right)-\delta\left(\sigma-\eta/2\right)\right)\\
= & -\frac{2\alpha}{r}\lim_{\epsilon\to0}\lim_{\eta\to0}\frac{1}{\eta}\left(\sin\left(-\eta r/2\right)-\sin\left(\eta r/2\right)\right)=\frac{4\alpha}{r}\lim_{\eta\to0}\frac{1}{\eta}\left(\sin\left(\eta r/2\right)\right)\\
= & 2\alpha
\end{flalign*}

Provided $\omega\to0$, of course.

\subsubsection{Summary}

Given the filter:
\begin{multline}
h\left(t-t',\,\left|\vec{r}-\vec{r}'\right|\right)=\frac{-i}{4\pi^{2}}\lim_{\epsilon\to0}\int_{-\frac{\epsilon}{2}}^{\frac{\epsilon}{2}}d\zeta\left(\zeta-\frac{\partial_{t}}{ic}\right)\cdot\\
\quad\quad\cdot\frac{\delta\left(\frac{\left|\vec{r}-\vec{r}'\right|}{c}-t+t'\right)\exp\left(i\zeta\left|\vec{r}-\vec{r}'\right|\right)-\delta\left(\frac{\left|\vec{r}-\vec{r}'\right|}{c}+t-t'\right)\exp\left(-i\zeta\left|\vec{r}-\vec{r}'\right|\right)}{\left|\vec{r}-\vec{r}'\right|}\label{eq:FullFilter}
\end{multline}

The action on plane wave is:
\begin{flalign*}
h\otimes\exp\left(i\left(\omega t-\vec{k}.\vec{r}\right)\right)= & \begin{cases}
\exp\left(i\left(\omega t-\vec{k}.\vec{r}\right)\right), & \omega\neq0,\,k>0,\,\omega=\pm ck\\
2, & k=0,\,\omega=0\\
0, & otherwise
\end{cases}\\
= & \exp\left(i\left(\omega t-\vec{k}.\vec{r}\right)\right)\cdot\,\left(\delta_{ck,\,\omega}+\delta_{-ck,\,\omega}\right),\,k\ge0
\end{flalign*}

\subsection{Green function and plane-wave decompositions}

Here we will reproduce the standard relation found, for example, in
Jackson~Eq.~(9.98) \cite{jackson}:
\begin{gather}
\frac{\exp\left(ik\left|\vec{r}-\vec{r}'\right|\right)}{\left|\vec{r}-\vec{r}'\right|}=i4\pi k\sum_{lm}j_{l}\left(kr_{<}\right)\left(j_{l}\left(kr_{>}\right)+in_{l}\left(kr_{>}\right)\right)Y_{lm}\left(\vec{\hat{r}}'\right)^{\dagger}Y_{lm}\left(\vec{\hat{r}}\right)\label{eq:GreenExpPosK}\\
\frac{\exp\left(-ik\left|\vec{r}-\vec{r}'\right|\right)}{\left|\vec{r}-\vec{r}'\right|}=-i4\pi k\sum_{lm}j_{l}\left(kr_{<}\right)\left(j_{l}\left(kr_{>}\right)-in_{l}\left(kr_{>}\right)\right)Y_{lm}\left(\vec{\hat{r}}'\right)Y_{lm}\left(\vec{\hat{r}}\right)^{\dagger}\label{eq:GreenExpNegK}
\end{gather}

Where $k\ge0$. It follows that:
\begin{flalign}
\frac{\sin\left(k\left|\vec{r}-\vec{r}'\right|\right)}{\left|\vec{r}-\vec{r}'\right|}= & \frac{1}{i2}i4\pi k\sum_{lm}j_{l}\left(kr_{<}\right)j_{l}\left(kr_{>}\right)\left(Y_{lm}\left(\vec{\hat{r}}'\right)^{\dagger}Y_{lm}\left(\vec{\hat{r}}\right)+Y_{lm}\left(\vec{\hat{r}}'\right)Y_{lm}\left(\vec{\hat{r}}\right)^{\dagger}\right)\nonumber \\
\frac{\sin\left(k\left|\vec{r}-\vec{r}'\right|\right)}{\left|\vec{r}-\vec{r}'\right|}= & 2\pi k\sum_{lm}j_{l}\left(kr\right)j_{l}\left(kr'\right)\left(Y_{lm}\left(\vec{\hat{r}}'\right)^{\dagger}Y_{lm}\left(\vec{\hat{r}}\right)+Y_{lm}\left(\vec{\hat{r}}'\right)Y_{lm}\left(\vec{\hat{r}}\right)^{\dagger}\right)\label{eq:GreenExpSin}
\end{flalign}

In case of $k=0$ we have (Jackson Eq.~(3.70) \cite{jackson}):
\begin{equation}
\frac{1}{\left|\vec{r}-\vec{r}'\right|}=4\pi\sum_{lm}\frac{1}{2l+1}\cdot\frac{r_{<}^{l}}{r_{>}^{l+1}}Y_{lm}\left(\vec{\hat{r}}'\right)^{\dagger}Y_{lm}\left(\vec{\hat{r}}\right)\label{eq:LapDec}
\end{equation}

Another useful expression from the same source is Eq.~(10.43) \cite{jackson}:
\begin{gather}
\exp\left(i\vec{k}.\vec{r}\right)=4\pi\sum_{lm}i^{l}j_{l}\left(kr\right)Y_{lm}\left(\vec{\hat{r}}\right)Y_{lm}\left(\vec{\hat{k}}\right)^{\dagger}\label{eq:PlaneWPos}\\
\exp\left(-i\vec{k}.\vec{r}\right)=4\pi\sum_{lm}\left(-i\right)^{l}j_{l}\left(kr\right)Y_{lm}\left(\vec{\hat{r}}\right)^{\dagger}Y_{lm}\left(\vec{\hat{k}}\right)\label{eq:PlaneWNeg}
\end{gather}

One can now substitute $\vec{\hat{k}}\to-\vec{\hat{k}}$ into Eq.~(\ref{eq:PlaneWNeg})
and use $Y_{lm}\left(-\vec{\hat{k}}\right)=\left(-1\right)^{l}Y_{lm}\left(\vec{\hat{k}}\right)$:
\begin{equation}
\exp\left(i\vec{k}.\vec{r}\right)=4\pi\sum_{lm}i^{l}j_{l}\left(kr\right)Y_{lm}\left(\vec{\hat{r}}\right)^{\dagger}Y_{lm}\left(\vec{\hat{k}}\right)\label{eq:PlaneWPosAlt}
\end{equation}

\subsection{Integral $\int_{0}^{\infty}dx\,\exp\left(\pm i\alpha x\right)$}

The integral can be evaluated by regularizing it and then taking the
limit:
\begin{flalign*}
I\left(\alpha\right) & =\int_{0}^{\infty}dx\,\exp\left(\pm i\alpha x\right)=\lim_{\nu\to0^{+}}\int_{0}^{\infty}dx\,\exp\left(\pm i\left(\alpha\pm i\nu\right)x\right)\\
 & =\lim_{\nu\to0^{+}}\frac{1}{\pm i\left(\alpha\pm i\nu\right)}\left(0-1\right)=\lim_{\nu\to0^{+}}\frac{\pm i}{\alpha\pm i\nu}
\end{flalign*}

Integrating $I\left(\alpha\right)$ with a good function we find:
\begin{equation}
\int_{0}^{\infty}dx\,\exp\left(\pm i\alpha x\right)=\text{PV}\frac{\pm i}{\alpha}+\pi\delta\left(\alpha\right)\label{eq:HalfExpInt}
\end{equation}

\subsection{Integrals $\int_{0}^{\infty}dx\,x\,\sin\sigma x$,~$\int_{0}^{\infty}dx\,x\,\cos\sigma x$ }

This integral can be evaluated in much the same way as more familiar
half-space exponential integrals:
\begin{flalign*}
\int_{0}^{\infty}dx\,x\,\exp\left(\pm i\sigma x\right)= & \lim_{\nu\to0}\int_{0}^{\infty}dx\,x\,\exp\left(\pm i\left(\sigma\pm i\nu\right)x\right)\\
q_{\pm}\left(\sigma\right)= & \lim_{\nu\to0}\frac{1}{\left(i\sigma\mp\nu\right)^{2}}\\
= & -\text{PV}\frac{1}{\sigma^{2}}+2\delta\left(\sigma\right)\lim_{a\to0}\frac{1}{a}\mp i\pi\delta'\left(\sigma\right)
\end{flalign*}

The last step has been obtained by evaluating the integral $\int d\sigma f\left(\sigma\right)q\left(\sigma\right)$
with a good function $f$. Allowing for sign of $\sigma$ to be positive
or negative we can (still) write:
\begin{equation}
\int_{0}^{\infty}dx\,x\,\exp\left(\pm i\sigma x\right)=-\text{PV}\frac{1}{\sigma^{2}}+2\delta\left(\sigma\right)\lim_{a\to0}\frac{1}{a}\mp i\pi\delta'\left(\sigma\right)\label{eq:xExpIX}
\end{equation}

It follows that:
\begin{flalign}
\int_{0}^{\infty}dx\,x\,\sin\sigma x= & -\pi\delta'\left(\sigma\right)\label{eq:xSinX}\\
\int_{0}^{\infty}dx\,x\,\cos\sigma x= & -\text{PV}\frac{1}{\sigma^{2}}+2\delta\left(\sigma\right)\lim_{a\to0}\frac{1}{a}\label{eq:xCosX}
\end{flalign}

\subsection{Integral $I_{\pm}\left(\sigma,\,\vec{r},\,\vec{k}\protect\neq0\right)$\label{app:Iint}}

We start from definition:
\[
I_{\pm}\left(\sigma,\,\vec{r},\,\vec{k}\neq0\right)=\int d^{3}r'\,\exp\left(\pm i\vec{k}.\vec{r}'\right)\cdot\frac{\sin\left(\sigma\left|\vec{r}-\vec{r}'\right|\right)}{\left|\vec{r}-\vec{r}'\right|}
\]

Clearly $I_{\pm}\left(-\sigma,\,\vec{r},\,\vec{k}\right)=-I_{\pm}\left(\sigma,\,\vec{r},\,\vec{k}\right)$
which suggests we do not need to worry about negative values of $\sigma$.
Formally $I_{\pm}\left(0,\,\vec{r},\,\vec{k}\neq0\right)=0$.

Expanding the relevant terms in spherical harmonics \cite{arfken2012mathematical}:

\begin{flalign*}
I_{+}&\left(\sigma>0,\,\vec{r},\,\vec{k}\neq0\right)= \\
&\quad= \int d^{3}r'\,\exp\left(i\vec{k}.\vec{r}'\right)\cdot\frac{\sin\left(\sigma\left|\vec{r}-\vec{r}'\right|\right)}{\left|\vec{r}-\vec{r}'\right|}\\
&\quad= \int d^{3}r'\,\left(4\pi\sum_{lm}i^{l}j_{l}\left(kr'\right)Y_{lm}\left(\vec{\hat{r}}'\right)Y_{lm}\left(\vec{\hat{k}}\right)^{\dagger}\right)\\
&\quad\quad \left(2\pi\sigma\sum_{l'm'}j_{l'}\left(\sigma r\right)j_{l'}\left(\sigma r'\right)\left(Y_{l'm'}\left(\vec{\hat{r}}'\right)^{\dagger}Y_{l'm'}\left(\vec{\hat{r}}\right)+Y_{l'm'}\left(\vec{\hat{r}}'\right)Y_{l'm'}\left(\vec{\hat{r}}\right)^{\dagger}\right)\right)\\
&\quad= \int_{0}^{\infty}dr'\,r'^{2}\left(4\pi\sum_{lm}i^{l}j_{l}\left(kr'\right)Y_{lm}\left(\vec{\hat{k}}\right)^{\dagger}\right)\\
 &\quad\quad \left(2\pi\sigma\sum_{l'm'}j_{l'}\left(\sigma r\right)j_{l'}\left(\sigma r'\right)\left(\delta_{ll'}\delta_{mm'}Y_{l'm'}\left(\vec{\hat{r}}\right)+\left(-1\right)^{m}\delta_{ll'}\delta_{m,-m'}Y_{l'm'}\left(\vec{\hat{r}}\right)^{\dagger}\right)\right)\\
&\quad= 8\pi^{2}\sigma\sum_{lm}i^{l}Y_{lm}\left(\vec{\hat{k}}\right)^{\dagger}\int_{0}^{\infty}dr'\,r'^{2}j_{l}\left(kr'\right)\cdot\left(j_{l}\left(\sigma r\right)j_{l}\left(\sigma r'\right)\left(Y_{lm}\left(\vec{\hat{r}}\right)+Y_{lm}\left(\vec{\hat{r}}\right)\right)\right)\\
&\quad= 16\pi^{2}\sigma\sum_{lm}i^{l}j_{l}\left(\sigma r\right)Y_{lm}\left(\vec{\hat{k}}\right)^{\dagger}Y_{lm}\left(\vec{\hat{r}}\right)\int_{0}^{\infty}dx\,x^{2}j_{l}\left(kx\right)\cdot j_{l}\left(\sigma x\right)
\end{flalign*}

One can then use a standard relation (e.g. Ref.~\cite{Maximon91}):
\[
\int_{0}^{\infty}dx\,x^{2}\,j_{l}\left(kx\right)j_{l}\left(\sigma x\right)=\frac{\pi}{2k\sigma}\delta\left(\sigma-k\right)=\frac{\pi}{2\sigma^{2}}\delta\left(\sigma-k\right)
\]

So:
\begin{flalign*}
I_{+}\left(\sigma>0,\,\vec{r},\,\vec{k}\neq0\right)= & 16\pi^{2}\sigma\sum_{lm}i^{l}j_{l}\left(\sigma r\right)Y_{lm}\left(\vec{\hat{k}}\right)^{\dagger}Y_{lm}\left(\vec{\hat{r}}\right)\frac{\pi}{2\sigma^{2}}\delta\left(\sigma-k\right)\\
= & \frac{8\pi^{3}}{\sigma}\delta\left(\sigma-k\right)\sum_{lm}i^{l}j_{l}\left(\sigma r\right)Y_{lm}\left(\vec{\hat{k}}\right)^{\dagger}Y_{lm}\left(\vec{\hat{r}}\right)\\
= & \frac{2\pi^{2}}{\sigma}\delta\left(\sigma-k\right)\cdot4\pi\sum_{lm}i^{l}j_{l}\left(kr\right)Y_{lm}\left(\vec{\hat{k}}\right)^{\dagger}Y_{lm}\left(\vec{\hat{r}}\right)\\
= & \frac{2\pi^{2}}{\sigma}\delta\left(\sigma-k\right)\cdot\exp\left(i\vec{k}.\vec{r}\right)
\end{flalign*}

From earlier considerations we also know that:
\[
I_{+}\left(\sigma<0,\vec{r},\,\vec{k}\neq0\right)=-I_{+}\left(-\sigma>0,\vec{r},\,\vec{k}\neq0\right)=\frac{2\pi^{2}}{\sigma}\delta\left(\sigma+k\right)\cdot\exp\left(i\vec{k}.\vec{r}\right)
\]

Given, $k>0$ one can cover both cases:
\[
I_{+}\left(\sigma\neq0,\vec{r},\,\vec{k}\neq0\right)=\frac{2\pi^{2}}{\sigma}\left(\delta\left(\sigma-k\right)+\delta\left(\sigma+k\right)\right)\cdot\exp\left(i\vec{k}.\vec{r}\right)
\]

\subsubsection{Full expression}

One can combine all the results and apply complex conjugation to get:
\begin{multline}
I_{\pm}\left(\sigma,\vec{r},\,\vec{k}\neq0\right)=\int d^{3}r'\,\exp\left(\pm i\vec{k}.\vec{r}'\right)\cdot\frac{\sin\left(\sigma\left|\vec{r}-\vec{r}'\right|\right)}{\left|\vec{r}-\vec{r}'\right|}=\\
\quad\quad=\begin{cases}
\frac{2\pi^{2}}{\sigma}\left(\delta\left(\sigma-k\right)+\delta\left(\sigma+k\right)\right)\cdot\exp\left(\pm i\vec{k}.\vec{r}\right), & \sigma\neq0\\
0, & \sigma=0
\end{cases}\label{eq:FullIInt}
\end{multline}

\subsection{Integral $I\left(\sigma,\,\vec{r}\right)$\label{app:IintK0}}

We start from definition:
\[
I\left(\sigma,\,\vec{r}\right)=\int d^{3}r'\frac{\sin\left(\sigma\left|\vec{r}-\vec{r}'\right|\right)}{\left|\vec{r}-\vec{r}'\right|}
\]

Clearly $I\left(-\sigma,\,\vec{r}\right)=-I\left(\sigma,\,\vec{r}\right)$
which suggests we do not need to worry about negative values of $\sigma$.
For $\sigma=0$ one formally gets $I\left(0,\,\vec{r}\right)=0$.
So we will focus on positive values. Using Eq.~(\ref{eq:GreenExpSin})
\begin{flalign*}
I\left(\sigma>0,\,\vec{r}\right)= & \int d^{3}r'\frac{\sin\left(\sigma\left|\vec{r}-\vec{r}'\right|\right)}{\left|\vec{r}-\vec{r}'\right|}\\
= & \int d^{3}r'2\pi\sigma\sum_{lm}j_{l}\left(\sigma r\right)j_{l}\left(\sigma r'\right)\left(Y_{lm}\left(\vec{\hat{r}}'\right)^{\dagger}Y_{lm}\left(\vec{\hat{r}}\right)+Y_{lm}\left(\vec{\hat{r}}'\right)Y_{lm}\left(\vec{\hat{r}}\right)^{\dagger}\right)\\
= & 2\pi\sigma j_{0}\left(\sigma r\right)\int_{0}^{\infty}dr'\,r'^{2}j_{0}\left(\sigma r'\right)\left(1+1\right)\\
= & 4\pi\,\frac{\sin\left(\sigma r\right)}{r\sigma}\int_{0}^{\infty}dx\,x\,\sin\left(\sigma x\right)
\end{flalign*}

Using Eq.~(\ref{eq:xSinX}):
\begin{flalign*}
I\left(\sigma>0,\,\vec{r}\right)= & 4\pi\,\frac{\sin\left(\sigma r\right)}{r\sigma}\left(-\pi\delta'\left(\sigma\right)\right)\\
= & -4\pi^{2}\frac{\sin\left(\sigma r\right)}{r\sigma}\delta'\left(\sigma\right)
\end{flalign*}

This is sufficient:
\begin{equation}
I\left(\sigma,\,\vec{r}\right)=\int d^{3}r'\frac{\sin\left(\sigma\left|\vec{r}-\vec{r}'\right|\right)}{\left|\vec{r}-\vec{r}'\right|}=\begin{cases}
-4\pi^{2}\frac{\sin\left(\sigma r\right)}{r\sigma}\delta'\left(\sigma\right), & \sigma\neq0\\
0, & otherwise
\end{cases}\label{eq:IintK0}
\end{equation}

\subsection{Apply to Flying Doughnut seed function\label{app:FDSeed}}

\subsubsection{Define the seed -function\label{sec:seedDef}}

We start from the seed function\cite{Hellwarth96}:
\begin{equation}
f\left(t,\,\vec{r}\right)=f\left(t,\,\rho,z\right)=\frac{1}{\rho^{2}-\left(\left(ct-z\right)+iq_{1}\right)\left(\left(ct+z\right)+iq_{2}\right)}\label{eq:seedDef}
\end{equation}

Can the denominator be zero for real-valued inputs with $0<q_{1}\le q_{2}$?
The imaginary part of the denominator is:
\[
\Im\left(\frac{1}{f\left(t,\,\rho,z\right)}\right)=\left(q_{2}-q_{1}\right)z-\left(q_{1}+q_{2}\right)ct
\]

We have $\Im\left(1/f\left(t,\,\rho,z\right)\right)=0$ if:
\begin{enumerate}
\item $q_{1}=q_{2}=q>0$ and $t=0$, the real part is then: $\Re\left(1/f\left(t,\,\rho,z\right)\right)\to q^{2}+z^{2}+\rho^{2}>0$
as long as $q>0$.
\item $q_{2}>q_{1}$ and $z=ct=0$, the real part is then: $\Re\left(1/f\left(t,\,\rho,z\right)\right)\to q_{1}q_{2}+\rho^{2}>0$
as long as $q_{1,2}>0$.
\item $q_{2}>q_{1}$ and $z=\left(\frac{q_{2}+q_{1}}{q_{2}-q_{1}}\right)ct$,
the real part is then: $\Re\left(1/f\left(t,\,\rho,z\right)\right)\to q_{1}q_{2}\left(1+\left(\frac{2ct}{q_{2}-q_{1}}\right)^{2}\right)+\rho^{2}>0$
as long as $q_{1,2}>0$.
\end{enumerate}
So either the real or the imaginary parts of the denominator are not
zero, thus $f\left(t,\,\rho,\,z\right)$ is well-behaved $\forall\,z,\,t,\,\rho,\,q_{1,2}\in\mathbb{R}$
if $q_{2}\ge q_{1}$ and $q_{1}>0$.

\subsubsection{Apply filter to seed function}

The filtered function is equal to (see Eq.~(\ref{eq:FullFilter})):
\begin{flalign}
F\left(t,\,\vec{r}\right)= & h\otimes f=\frac{-i}{4\pi^{2}}\lim_{\epsilon\to0}\int_{-\frac{\epsilon}{2}}^{\frac{\epsilon}{2}}d\zeta\left(\zeta-\frac{\partial_{t}}{ic}\right)\int dt'\int d^{3}r'\,f\left(t',\,\vec{r}'\right)\nonumber \\
 & \cdot\left(\frac{\delta\left(\frac{\left|\vec{r}-\vec{r}'\right|}{c}-t+t'\right)\exp\left(i\zeta\left|\vec{r}-\vec{r}'\right|\right)-\delta\left(\frac{\left|\vec{r}-\vec{r}'\right|}{c}+t-t'\right)\exp\left(-i\zeta\left|\vec{r}-\vec{r}'\right|\right)}{\left|\vec{r}-\vec{r}'\right|}\right)\nonumber \\
= & \frac{-i}{4\pi^{2}}\lim_{\epsilon\to0}\int_{-\frac{\epsilon}{2}}^{\frac{\epsilon}{2}}d\zeta\left(\zeta-\frac{\partial_{t}}{ic}\right)\int d^{3}r'\,\nonumber \\
 & \cdot\left(\frac{f\left(t-\frac{\left|\vec{r}-\vec{r}'\right|}{c},\,\vec{r}'\right)\exp\left(i\zeta\left|\vec{r}-\vec{r}'\right|\right)-f\left(t+\frac{\left|\vec{r}-\vec{r}'\right|}{c},\,\vec{r}'\right)\exp\left(-i\zeta\left|\vec{r}-\vec{r}'\right|\right)}{\left|\vec{r}-\vec{r}'\right|}\right)\nonumber \\
= & \frac{-i}{4\pi^{2}}\lim_{\epsilon\to0}\int_{-\frac{\epsilon}{2}}^{\frac{\epsilon}{2}}d\zeta\left(\zeta-\frac{\partial_{t}}{ic}\right)\oint d^{2}\Omega'\nonumber \\
 & \cdot\Biggl[\int_{0}^{\infty}dr'\cdot\exp\left(i\zeta\left|\vec{r}-\vec{r}'\right|\right)\cdot\frac{r'}{\left|\vec{r}-\vec{r}'\right|}\cdot r'f\left(t-\frac{\left|\vec{r}-\vec{r}'\right|}{c},\,\vec{r}'\right)+\nonumber \\
 & +\int_{0}^{\infty}dr'\cdot\exp\left(-i\zeta\left|\vec{r}-\vec{r}'\right|\right)\cdot\frac{-r'}{\left|\vec{r}-\vec{r}'\right|}\cdot r'f\left(t+\frac{\left|\vec{r}-\vec{r}'\right|}{c},\,\vec{r}'\right)\Biggr]\label{eq:bigFStart}
\end{flalign}

Define:
\begin{equation}
p_{\pm}=\int_{0}^{\infty}dr'\cdot\exp\left(\pm i\zeta\left|\vec{r}-\vec{r}'\right|\right)\cdot\frac{\pm r'}{\left|\vec{r}-\vec{r}'\right|}\cdot r'f\left(t\mp\frac{\left|\vec{r}-\vec{r}'\right|}{c},\,\vec{r}'\right)\label{eq:pDef}
\end{equation}

The key thing we want from $p_{\pm}$ is some sort of delta function
because anything finite will be removed by the $\zeta$-integral.
Due to arguments in Sec.~\ref{sec:seedDef} it is clear that the
integrand is well-behaved for all finite $r'$, so we can only pick
up $\delta$ at infinity. We can therefore replace the integrand
with its $r'\to\infty$ approximation. As long as the new integrand
behaves well at finite $r'$ we only introduce finite error (which
will be removed by $\zeta$-integral). Let us expand the non-exponential
factors at large $r'$.

The basic fraction:
\begin{equation}
\lim_{r'\to\infty}\frac{\pm r'}{\left|\vec{r}-\vec{r}'\right|}=\pm1\mp\frac{\vec{r}.\vec{r}'}{r'^{2}}=\pm1+\mathcal{O}\left(\frac{1}{r'}\right)\label{eq:SimpFracLim}
\end{equation}

The seed function part. For now, it is more convenient to look at:
\begin{flalign*}
\lim_{r'\to\infty}&\left(r'f\left(t\mp\frac{\left|\vec{r}-\vec{r}'\right|}{c},\,\vec{r}'\right)\right)^{-1}= \\
&\quad\quad= \lim_{r'\to\infty}r'^{-1}\left[\rho'^{2}-\left(\left(ct\mp\left|\vec{r}-\vec{r}'\right|-z'\right)+iq_{1}\right)\left(\left(ct\mp\left|\vec{r}-\vec{r}'\right|+z'\right)+iq_{2}\right)\right]\\
&\quad\quad= \lim_{r'\to\infty}r'^{-1}\Biggl[\rho'^{2}-\left(ct\mp\left|\vec{r}-\vec{r}'\right|-z'\right)\cdot\left(ct\mp\left|\vec{r}-\vec{r}'\right|+z'\right)-iq_{2}\left(ct\mp\left|\vec{r}-\vec{r}'\right|-z'\right)-\\
&\quad\quad\quad -iq_{1}\left(ct\mp\left|\vec{r}-\vec{r}'\right|+z'\right)+q_{1}q_{2}\Biggr]\\
&\quad\quad= \lim_{r'\to\infty}r'^{-1}\Biggl[\rho'^{2}-\left(ct\mp\left|\vec{r}-\vec{r}'\right|\right)^{2}+z'^{2}-iq_{2}\left(ct\mp\left|\vec{r}-\vec{r}'\right|-z'\right)-\\
&\quad\quad\quad -iq_{1}\left(ct\mp\left|\vec{r}-\vec{r}'\right|+z'\right)+q_{1}q_{2}\Biggr]\\
&\quad\quad= \lim_{r'\to\infty}r'^{-1}\Biggl[r'^{2}-\left(ct\right)^{2}\pm2ct\left|\vec{r}-\vec{r}'\right|-r^{2}-r'^{2}+2\vec{r}.\vec{r}'-iq_{2}\left(ct\mp\left|\vec{r}-\vec{r}'\right|-z'\right)-\\
&\quad\quad\quad -iq_{1}\left(ct\mp\left|\vec{r}-\vec{r}'\right|+z'\right)+q_{1}q_{2}\Biggr]\\
&\quad\quad= \lim_{r'\to\infty}r'^{-1}\Biggl[-\left(ct\right)^{2}\pm2ct\left|\vec{r}-\vec{r}'\right|-r^{2}+2\vec{r}.\vec{r}'-iq_{2}\left(ct\mp\left|\vec{r}-\vec{r}'\right|-z'\right)-\\
&\quad\quad\quad -iq_{1}\left(ct\mp\left|\vec{r}-\vec{r}'\right|+z'\right)+q_{1}q_{2}\Biggr]
\end{flalign*}

The inside of the bracket can now be expanded up to $\mathcal{O}\left(r'^{0}\right)$:
\begin{flalign*}
\lim_{r'\to\infty}&\left(r'f\left(t\mp\frac{\left|\vec{r}-\vec{r}'\right|}{c},\,\vec{r}'\right)\right)^{-1}=\\ 
&\quad\quad= \lim_{r'\to\infty}r'^{-1}\Biggl[\pm2ctr'+2\vec{r}.\vec{r}'-iq_{2}\left(\mp r'-r'\left(\vec{\hat{z}}.\vec{\hat{r}}'\right)\right)-\\
&\quad\quad\quad -iq_{1}\left(\mp r'+r'\left(\vec{\hat{z}}.\vec{\hat{r}}'\right)\right)+\mathcal{O}\left(r'^{0}\right)\Biggr]\\
&\quad\quad= \pm2ct+2\vec{r}.\vec{\hat{r}}'\pm iq_{2}+iq_{2}\left(\vec{\hat{z}}.\vec{\hat{r}}'\right)\pm iq_{1}-iq_{1}\left(\vec{\hat{z}}.\vec{\hat{r}}'\right)+\mathcal{O}\left(r'^{-1}\right)\\
&\quad\quad= \left(2\vec{r}.\vec{\hat{r}}'+i\left(\vec{\hat{z}}.\vec{\hat{r}}'\right)\left(q_{2}-q_{1}\right)\right)\pm\left(2ct+i\left(q_{2}+q_{1}\right)\right)+\mathcal{O}\left(r'^{-1}\right)
\end{flalign*}

Therefore:
\begin{equation}
\lim_{r'\to\infty}\left(r'f\left(t\mp\frac{\left|\vec{r}-\vec{r}'\right|}{c},\,\vec{r}'\right)\right)^{-1}=\frac{1}{\left[2\vec{r}.\vec{\hat{r}}'+i\left(\vec{\hat{z}}.\vec{\hat{r}}'\right)\left(q_{2}-q_{1}\right)\right]\pm\left[2ct+i\left(q_{2}+q_{1}\right)\right]}+\mathcal{O}\left(\frac{1}{r'}\right)\label{eq:fLim}
\end{equation}

It follows that the integrand of Eq.~(\ref{eq:pDef}) can be thought
of as:
\begin{multline*}
\exp\left(\pm i\zeta\left|\vec{r}-\vec{r}'\right|\right)\cdot\frac{\pm r'}{\left|\vec{r}-\vec{r}'\right|}\cdot r'f\left(t\mp\frac{\left|\vec{r}-\vec{r}'\right|}{c},\,\vec{r}'\right)=\\
\exp\left(\mp i\zeta\vec{\hat{r}}'.\vec{r}\right)\exp\left(\pm i\zeta r'\right)\left(\frac{\pm}{\left[2\vec{r}.\vec{\hat{r}}'+i\left(\vec{\hat{z}}.\vec{\hat{r}}'\right)\left(q_{2}-q_{1}\right)\right]\pm\left[2ct+i\left(q_{2}+q_{1}\right)\right]}+\mathcal{H}\left(r'\right)\right)
\end{multline*}

Function $\mathcal{H}$ is finite at all times, at $r'\to0$ the function
tends to the value $\left(-1\right)\times\left(first\,term\right)$
(to make the full expression go to zero). At large $r'$, $\mathcal{H}\propto\frac{1}{r'}$.
Since
\[
\int_{1}^{\infty}\frac{\exp\left(ix\right)}{x}=\Gamma\left(0,-i\right)=-0.337404+i0.624713=finite
\]

we can say:
\[
p_{\pm}=\frac{\pm\exp\left(\mp i\zeta\vec{\hat{r}}'.\vec{r}\right)}{\left[2\vec{r}.\vec{\hat{r}}'+i\left(\vec{\hat{z}}.\vec{\hat{r}}'\right)\left(q_{2}-q_{1}\right)\right]\pm\left[2ct+i\left(q_{2}+q_{1}\right)\right]}\cdot\int_{0}^{\infty}dx\,\exp\left(\pm i\zeta x\right)+finite
\]

Using Eq.~(\ref{eq:HalfExpInt}):
\[
p_{\pm}=\frac{\pm\exp\left(\mp i\zeta\vec{\hat{r}}'.\vec{r}\right)}{\left[2\vec{r}.\vec{\hat{r}}'+i\left(\vec{\hat{z}}.\vec{\hat{r}}'\right)\left(q_{2}-q_{1}\right)\right]\pm\left[2ct+i\left(q_{2}+q_{1}\right)\right]}\cdot\left(\text{PV}\frac{\pm i}{\zeta}+\pi\delta\left(\zeta\right)\right)+finite
\]

Coming back to Eq.~(\ref{eq:bigFStart}):

\begin{flalign*}
F\left(t,\,\vec{r}\right)= & \frac{-i}{4\pi^{2}}\lim_{\epsilon\to0}\int_{-\frac{\epsilon}{2}}^{\frac{\epsilon}{2}}d\zeta\left(\zeta-\frac{\partial_{t}}{ic}\right)\oint d^{2}\Omega'\left[p_{+}+p_{-}\right]\\
= & \frac{-i}{4\pi^{2}}\oint d^{2}\Omega'\lim_{\epsilon\to0}\int_{-\frac{\epsilon}{2}}^{\frac{\epsilon}{2}}d\zeta\left(\zeta-\frac{\partial_{t}}{ic}\right)\left[p_{+}+p_{-}\right]\\
= & \frac{-i}{4\pi^{2}}\oint d^{2}\Omega'\Biggl[\frac{Q_{+}^{\left(1\right)}}{\left[2\vec{r}.\vec{\hat{r}}'+i\left(\vec{\hat{z}}.\vec{\hat{r}}'\right)\left(q_{2}-q_{1}\right)\right]+\left[2ct+i\left(q_{2}+q_{1}\right)\right]}+\\
 & +\frac{Q_{-}^{\left(1\right)}}{\left[2\vec{r}.\vec{\hat{r}}'+i\left(\vec{\hat{z}}.\vec{\hat{r}}'\right)\left(q_{2}-q_{1}\right)\right]-\left[2ct+i\left(q_{2}+q_{1}\right)\right]}+\\
 & +\left(-\frac{\partial_{t}}{ic}\right)\frac{Q_{+}^{\left(0\right)}}{\left[2\vec{r}.\vec{\hat{r}}'+i\left(\vec{\hat{z}}.\vec{\hat{r}}'\right)\left(q_{2}-q_{1}\right)\right]+\left[2ct+i\left(q_{2}+q_{1}\right)\right]}+\\
 & +\left(-\frac{\partial_{t}}{ic}\right)\frac{Q_{-}^{\left(0\right)}}{\left[2\vec{r}.\vec{\hat{r}}'+i\left(\vec{\hat{z}}.\vec{\hat{r}}'\right)\left(q_{2}-q_{1}\right)\right]-\left[2ct+i\left(q_{2}+q_{1}\right)\right]}\Biggr]
\end{flalign*}

Consider the $\zeta$-integrals:
\begin{flalign*}
Q_{\pm}^{\left(1\right)}= & \pm\lim_{\epsilon\to0}\int_{-\frac{\epsilon}{2}}^{\frac{\epsilon}{2}}d\zeta\cdot\zeta\cdot\exp\left(\mp i\zeta\vec{\hat{r}}'.\vec{r}\right)\cdot\left(\text{PV}\frac{\pm i}{\zeta}+\pi\delta\left(\zeta\right)\right)\\
= & \pm\lim_{\epsilon\to0}\left(\pm i\int_{-\frac{\epsilon}{2}}^{\frac{\epsilon}{2}}d\zeta\cdot\exp\left(\mp i\zeta\vec{\hat{r}}'.\vec{r}\right)+\pi\cdot0\cdot\exp\left(\mp i\left(0\right)\vec{\hat{r}}'.\vec{r}\right)\right)\\
= & i\lim_{\epsilon\to0}\int_{-\frac{\epsilon}{2}}^{\frac{\epsilon}{2}}d\zeta\cdot\exp\left(\mp i\zeta\vec{\hat{r}}'.\vec{r}\right)=i\lim_{\epsilon\to0}\left(\frac{1}{\mp i\vec{\hat{r}}'.\vec{r}}\right)\left(\pm i2\sin\left(\epsilon\vec{\hat{r}}'.\vec{r}\right)\right)=0
\end{flalign*}

Note that this works even if $\vec{r}\to0$. The other integral:
\begin{flalign*}
Q_{\pm}^{\left(0\right)}= & \pm\lim_{\epsilon\to0}\int_{-\frac{\epsilon}{2}}^{\frac{\epsilon}{2}}d\zeta\cdot\exp\left(\mp i\zeta\vec{\hat{r}}'.\vec{r}\right)\cdot\left(\text{PV}\frac{\pm i}{\zeta}+\pi\delta\left(\zeta\right)\right)\\
= & \pm\left(\pm i\lim_{\epsilon\to0}\text{PV}\int_{-\frac{\epsilon}{2}}^{\frac{\epsilon}{2}}d\zeta\cdot\frac{\exp\left(\mp i\zeta\vec{\hat{r}}'.\vec{r}\right)}{\zeta}+\pi\right)\\
= & \pm\left(\pm i\lim_{\epsilon\to0}\left(\pm i2\text{Si}\left(\frac{\epsilon}{2}\right)\right)+\pi\right)\\
= & \pm\pi
\end{flalign*}

Where Si(...) is the sine-integral, that evaluates to zero at zero
argument. Therefore:
\begin{flalign*}
F\left(t,\,\vec{r}\right)= & \frac{-i}{4\pi^{2}}\lim_{\epsilon\to0}\int_{-\frac{\epsilon}{2}}^{\frac{\epsilon}{2}}d\zeta\left(\zeta-\frac{\partial_{t}}{ic}\right)\oint d^{2}\Omega'\left[p_{+}+p_{-}\right]\\
= & \frac{-i}{4\pi^{2}}\oint d^{2}\Omega'\lim_{\epsilon\to0}\int_{-\frac{\epsilon}{2}}^{\frac{\epsilon}{2}}d\zeta\left(\zeta-\frac{\partial_{t}}{ic}\right)\left[p_{+}+p_{-}\right]\\
= & \frac{-i}{4\pi^{2}}\oint d^{2}\Omega'\left(-\frac{\partial_{t}}{ic}\right)\Biggl[\frac{+\pi}{\left[2\vec{r}.\vec{\hat{r}}'+i\left(\vec{\hat{z}}.\vec{\hat{r}}'\right)\left(q_{2}-q_{1}\right)\right]+\left[2ct+i\left(q_{2}+q_{1}\right)\right]}+\\
 & +\frac{-\pi}{\left[2\vec{r}.\vec{\hat{r}}'+i\left(\vec{\hat{z}}.\vec{\hat{r}}'\right)\left(q_{2}-q_{1}\right)\right]-\left[2ct+i\left(q_{2}+q_{1}\right)\right]}\Biggr]\\
= & \frac{\partial_{t}}{4\pi c}\oint d^{2}\Omega'\left\{ \frac{-2\left[2ct+i\left(q_{2}+q_{1}\right)\right]}{\left[2\vec{r}.\vec{\hat{r}}'+i\left(\vec{\hat{z}}.\vec{\hat{r}}'\right)\left(q_{2}-q_{1}\right)\right]^{2}-\left[2ct+i\left(q_{2}+q_{1}\right)\right]^{2}}\right\} \\
= & -\frac{1}{2\pi c}\oint d^{2}\Omega'\partial_{t}\left\{ \frac{\left[2ct+i\left(q_{2}+q_{1}\right)\right]}{\left[2\vec{r}.\vec{\hat{r}}'+i\left(\vec{\hat{z}}.\vec{\hat{r}}'\right)\left(q_{2}-q_{1}\right)\right]^{2}-\left[2ct+i\left(q_{2}+q_{1}\right)\right]^{2}}\right\}
\end{flalign*}

Finally, we differentiate to get:
\begin{flalign*}
&F\left(t,\,\vec{r}\right)= \\
&\quad= -\frac{1}{2\pi c}\oint d^{2}\Omega'\partial_{t}\left\{ \frac{\left[2ct+i\left(q_{2}+q_{1}\right)\right]}{\left[2\vec{r}.\vec{\hat{r}}'+i\left(\vec{\hat{z}}.\vec{\hat{r}}'\right)\left(q_{2}-q_{1}\right)\right]^{2}-\left[2ct+i\left(q_{2}+q_{1}\right)\right]^{2}}\right\} \\
&\quad= -\frac{1}{2\pi c}\oint d^{2}\Omega'\cdot\\
&\quad\quad \cdot\left\{ \frac{\left(2c\right)\left(\left[2\vec{r}.\vec{\hat{r}}'+i\left(\vec{\hat{z}}.\vec{\hat{r}}'\right)\left(q_{2}-q_{1}\right)\right]^{2}-\left[2ct+i\left(q_{2}+q_{1}\right)\right]^{2}\right)}{\left(\left[2\vec{r}.\vec{\hat{r}}'+i\left(\vec{\hat{z}}.\vec{\hat{r}}'\right)\left(q_{2}-q_{1}\right)\right]^{2}-\left[2ct+i\left(q_{2}+q_{1}\right)\right]^{2}\right)^{2}}\:-\right. \\
&\quad\quad\quad\quad \left. -\:\frac{\left(2ct+i\left(q_{2}+q_{1}\right)\right)\left(-2\left[2ct+i\left(q_{2}+q_{1}\right)\right]\left(2c\right)\right)}{\left(\left[2\vec{r}.\vec{\hat{r}}'+i\left(\vec{\hat{z}}.\vec{\hat{r}}'\right)\left(q_{2}-q_{1}\right)\right]^{2}-\left[2ct+i\left(q_{2}+q_{1}\right)\right]^{2}\right)^{2}}\right\} \\
&\quad= -\frac{1}{\pi}\oint d^{2}\Omega'\cdot\\
&\quad\quad \cdot\left\{ \frac{\left[2\vec{r}.\vec{\hat{r}}'+i\left(\vec{\hat{z}}.\vec{\hat{r}}'\right)\left(q_{2}-q_{1}\right)\right]^{2}-\left[2ct+i\left(q_{2}+q_{1}\right)\right]^{2}+2\left(2ct+i\left(q_{2}+q_{1}\right)\right)^{2}}{\left(\left[2\vec{r}.\vec{\hat{r}}'+i\left(\vec{\hat{z}}.\vec{\hat{r}}'\right)\left(q_{2}-q_{1}\right)\right]^{2}-\left[2ct+i\left(q_{2}+q_{1}\right)\right]^{2}\right)^{2}}\right\} \\
&\quad= -\frac{1}{\pi}\oint d^{2}\Omega'\cdot\left\{ \frac{\left[2\vec{r}.\vec{\hat{r}}'+i\left(\vec{\hat{z}}.\vec{\hat{r}}'\right)\left(q_{2}-q_{1}\right)\right]^{2}+\left[2ct+i\left(q_{2}+q_{1}\right)\right]^{2}}{\left(\left[2\vec{r}.\vec{\hat{r}}'+i\left(\vec{\hat{z}}.\vec{\hat{r}}'\right)\left(q_{2}-q_{1}\right)\right]^{2}-\left[2ct+i\left(q_{2}+q_{1}\right)\right]^{2}\right)^{2}}\right\}
\end{flalign*}

So the filtered seed-function is:
\begin{equation}
F\left(t,\,\vec{r}\right)=-\frac{1}{\pi}\oint d^{2}\Omega'\cdot\left\{ \frac{\left[2\vec{r}.\vec{\hat{r}}'+i\left(\vec{\hat{z}}.\vec{\hat{r}}'\right)\left(q_{2}-q_{1}\right)\right]^{2}+\left[2ct+i\left(q_{2}+q_{1}\right)\right]^{2}}{\left(\left[2\vec{r}.\vec{\hat{r}}'+i\left(\vec{\hat{z}}.\vec{\hat{r}}'\right)\left(q_{2}-q_{1}\right)\right]^{2}-\left[2ct+i\left(q_{2}+q_{1}\right)\right]^{2}\right)^{2}}\right\} \label{eq:FilteredSeedSol}
\end{equation}

The integral over the solid angle can be left as it is, since it is
well-suited for numerical evaluation.

\bibliography{plane_wave_paper}

\end{document}